\documentclass[11pt,a4paper]{article}

\usepackage{jheppub}
\usepackage[usenames,dvipsnames]{xcolor}
\usepackage{blindtext}

\usepackage{amssymb} 
\usepackage{amsmath}
\usepackage{mathtools}
\usepackage{amsfonts}    
\usepackage{dsfont}
\usepackage{pdfpages}
\usepackage{verbatim}
\usepackage{stmaryrd}
\usepackage{graphicx}
\graphicspath{{images/}}
\usepackage{braket}
\usepackage[dvipsnames]{xcolor}

\usepackage{tikz}
\usetikzlibrary{arrows}
\usetikzlibrary{decorations.pathmorphing}
\usetikzlibrary{decorations.markings}
\usetikzlibrary{shapes.misc}
\tikzset{cross/.style={cross out, draw=black, minimum size=2*(#1-\pgflinewidth), inner sep=0pt, outer sep=0pt},
cross/.default={1pt}}
\usetikzlibrary{shapes.geometric}
\tikzset{
    partial ellipse/.style args={#1:#2:#3}{
        insert path={+ (#1:#3) arc (#1:#2:#3)}
    }
}

\renewcommand{\figurename}{Fig.}

\newcommand{\dd}{\text{d}}

\newcommand{\bi}{\begin{itemize}}
\newcommand{\ei}{\end{itemize}}
\newcommand{\bea}{\begin{eqnarray}}
\newcommand{\eea}{\end{eqnarray}}
\newcommand{\be}{\begin{equation}}
\newcommand{\ee}{\end{equation}}

 \definecolor{pink}{rgb}{1.0, 0.13, 0.32}

\numberwithin{equation}{section}
\allowdisplaybreaks  % allow page breaks in displayed eqs

\usepackage{subfiles}

\begin{document}
\vspace*{2.5cm}
\begin{center}
{ \LARGE \textsc{Aspects of Closed Matricial Worlds}\\}
\vspace*{2cm}

Dionysios Anninos$^{1,2}$ and Samuel Brian$^1$\\ 

\vspace*{0.5cm}

\vskip4mm
{
\footnotesize
{$^1$Department of Mathematics, King's College London, Strand, London WC2R 2LS, UK \newline
$^2$Instituut voor Theoretische Fysica, KU Leuven, Celestijnenlaan 200D, B-3001 Leuven, Belgium \newline
\\
}}

{\footnotesize{dionysios.anninos@kcl.ac.uk, \,\,  samuel.brian@kcl.ac.uk} } 
\end{center}

\vspace{1.5cm}

\noindent
\begin{abstract}
\noindent  We investigate the Hilbert space structure for simple theories of gravity with $\Lambda>0$ on closed spatial sections. Our motivation ties to the presence of  gravitational saddles in four-dimensional $\Lambda>0$ Einstein-Maxwell theory with $S^2\times \Sigma_h$ topology,  where $\Sigma_h$ is a genus-$h$ Riemann surface.  Here, as a concrete starting point, the problem is explored for two-dimensional $\Lambda>0$ quantum gravity. We revisit and elaborate on exact results in the matrix model literature. We study gravitational wavefunctions from both the perspective of the Wheeler-DeWitt equation and the gravitational path integral. Though simple, the setting displays many features of general interest such as large volume effects that disrupt the perturbative expansion, topological corrections to the path-integral wavefunction that offend the exact Wheeler-DeWitt equation, and a sphere path integral $\mathcal{Z}^{(0)}_{\text{grav}}$ with non-trivial structure in $\Lambda$. We discuss candidate inner products for the infinite-dimensional canonical gravitational Hilbert space uncovered by Lian and Zuckerman. By establishing explicit results up to genus-six, we argue that the dominant contribution to the genus-$h$ gravitational disk path integral at large spatial size mimics the behavior of two-dimensional topological gravity. In passing, we show that for $\mathcal{Z}^{(0)}_{\text{grav}}$ to give rise to a positive counting problem for discretised Riemann surfaces, it must have a negative pre-factor. We contrast our analysis to the more realistic case of timelike Liouville theory.
\end{abstract}

\vspace{1.4cm}

\begin{center}\textit{Dedicated to the memory of Umut G\"ursoy}\end{center}

\newpage

\tableofcontents

\section{Introduction}

The purpose of this article is to study the Hilbert space and associated wavefunctions for theories of gravity with a closed spatial section in a controlled setting. In the semiclassical regime, one meaning of a gravitational wavefunction is that it solves the Wheeler-DeWitt equation \cite{dewitt1967quantum}. Another possible meaning is that a gravitational wavefunction is the output of a gravitational path integral. An example of the latter is the no-boundary wavefunction of Hartle and Hawking \cite{hartle1983wave}, $\Psi_{\text{HH}}$, which is computed by a gravitational path integral in Euclidean signature. The path integrals are ordinarily computed in a saddle point approximation, or employing minisuperspace methods. Beyond such approximations, there is little understanding on what quantum effects gravitational wavefunctions might exhibit, and whether such effects can accumulate over sufficiently long time periods \cite{Guth:2007ng}. 
\newline\newline
In simple enough settings, $\Psi_{\text{HH}}$ can be argued to solve the Wheeler-DeWitt equation. However, the two meanings are generally distinct as $\Psi_{\text{HH}}$ is sensitive to a choice of contour for the fields that are path-integrated. For instance, a Euclidean gravitational path integral over a lapse function taken to be non-negative can yield outputs that violate the Hamiltonian constraint \cite{Teitelboim:1981ua,Teitelboim:1983fh}. Such violations may be less sensitive when the Hartle-Hawking path integral is taken on a manifold with a single boundary filled with a trivial topology. (A canonical example being an $S^3$ boundary filled in with a portion of $S^4$.) Beyond trivial topology, gravitational theories may admit a host of smooth Euclidean solutions, sometimes referred to as gravitational instantons \cite{Hawking:1976jb,eguchi1980gravitation}, displaying distinctive geometric and topological features. For an ordinary gauge theory, Euclidean instantons are directly related to the structure of the ground state wavefunction \cite{Vainshtein:1981wh}. In the gravitational case, partly due to the more enigmatic nature of Euclidean gravity, it is not entirely clear how gravitational instantons are coded into the  wavefunctions.\footnote{That is not to say their potential physical consequences have not been discussed in the literature. Indeed, it was pointed out in \cite{Kimura:1969iwz,Delbourgo:1972xb,Eguchi:1976db} that gravitational instantons with non-trivial Pontryagin density can contribute to the chiral anomaly.} 
\newline\newline
To make the discussion concrete, we will consider a gravitational theory with $\Lambda>0$ on a closed spatial slice, $\Sigma$. In four spacetime dimensions, even the pure gravitational theory displays a variety of gravitational instantons including $S^4$, $\mathbb{C}P^2$, and $S^2\times S^2$, among others. These were recently explored in \cite{Anninos:2025ltd}.  In the Hartle-Hawking literature, it is customary to take $\Sigma$ to have $S^3$ topology, and it was argued by Hartle and Hawking that a candidate ground state wavefunction is computed by path integrating the metric field on a compact manifold with a single $S^3$ boundary. The simplest instance of this is to fill the $S^3$ with a portion of $S^4$. However, we could also fill the $S^3$ with a portion of $\mathbb{C}P^2$ as this also admits $S^3$ slices. By the logic of Hartle and Hawking, this other filling of $S^3$ may  yield a contribution to $\Psi_{\text{HH}}$ provided the additional matter fields can be placed on such a topology. If we consider a slight extension to the Einstein-Maxwell theory with $\Lambda>0$, the story becomes richer. Here, one has a host of additional gravitational instantons admitting a topology $\Sigma_h \times S^2$ where $\Sigma_h$ is an arbitrary genus-$h$ closed Riemann surface \cite{Anninos:2025ltd}. We can now cut the $\Sigma_h$ along an $S^1$ to produce a potential contribution for $\Psi_{\text{HH}}$ on an $S^1 \times S^2$ spatial slice. The gravitational saddle fills the $S^1$ with a  Euclidean hyperbolic metric on a genus-$h$ surface. So in this case, we have an infinite set of potential contributions to the Hartle-Hawking wavefunction on an $S^1 \times S^2$ spatial slice. Moreover, we must integrate over the respective moduli space, $\mathcal{M}_h$, of the cut surface. Though the setup is four-dimensional, these particular contributions to the $S^1 \times S^2$ wavefunction have a two-dimensional flavour reminiscent of classic calculations in the context of two dimensional quantum gravity (as reviewed in \cite{Ginsparg:1993is}).\footnote{Indeed, an effective two-dimensional picture may yield correct results in certain regimes, such as the limit where the ratio of the $S^1$ to $S^2$ size is large (see for example \cite{Turiaci:2025xwi}), or in the presence of judiciously tuned magnetic fluxes, though such a picture generally requires a delicate treatment and we will treat it elsewhere \cite{EML}.}
\newline\newline
As a warm-up to the four-dimensional Einstein-Maxwell $\Lambda>0$ problem, we are thus motivated to  revisit and analyse cosmological wavefunctions in two-dimensional quantum gravity with $\Lambda>0$ on a spatial $S^1$. There is foundational work on this problem \cite{Banks:1989df,Moore:1991ir,Ambjorn:1990ji,Moore:1991sf}, as well as more recent work \cite{Betzios:2020nry,Betzios:2025eev}.  Our modest aim is to complement the existing results with an  emphasis on  the  Hilbert space viewpoint. We will consider both the gravitational path integral and Wheeler-DeWitt perspectives, as well as the complementary definition of these in terms of a double scaling limit of a matrix integral. Due to the latter connection, one could refer to these theories as closed matricial worlds.
\newline\newline
The simplicity of the two-dimensional problem allows us to compute the gravitational path integral on a Riemann surface of arbitrary genus. We can thus explicitly see how non-trivial topologies lead to violations of the two-dimensional Wheeler-DeWitt equation. What is more, being ultraviolet finite and renormalisable, the Wheeler-DeWitt equation for these theories is in much better shape than the four-dimensional one. Through its connection to Liouville theory, one can offer a sensible operator ordering prescription for the Wheeler-DeWitt equation \cite{Teitelboim:1983ux}, and there are no physically meaningful higher-derivative corrections that modify it. Though most of our presentation is dedicated to the simplest case of pure two-dimensional quantum gravity, the bulk of our discussion will also apply to the case where it is coupled to an arbitrary minimal model. Thus, in this simplified context, one can examine otherwise intractable features.
\subsection*{Outline of the paper}
The paper is structured in a way that meshes classic material with novel results. The reason for this structure is to streamline many important, but somewhat disjoint, results from the previous literature. We will specify what is modestly novel along the way. Section \ref{The Continuum Approach at genus-zero} discusses two-dimensional quantum gravity with $\Lambda>0$ in broad terms, and offers an effective field theoretic perspective on its simplicity. An emphasis is placed on both manifolds with and without boundaries. Section \ref{2dLiouville} reviews the description of two-dimensional gravity in terms of a Liouville quantum field theory. Although section \ref{2dLiouville}  is mostly review material, we do emphasise in section \ref{signZ} that the non-analytic part of the gravitational sphere path integral is negative if we are to relate it to a positive counting of discretised Riemann surfaces. In section \ref{LZsec}, we discus the infinite BRST cohomology of Lian and Zuckerman, and candidate inner products for this space of states. We discuss an inner-product for the Liouville sector that originates from the DOZZ formula. In section \ref{Wavefunctional perspective}, we review the construction from both a Wheeler-DeWitt and gravitational path-integral perspective. In section \ref{hgwf}, we compute the genus-$h$ corrections to the gravitational path integral for $h=1,2,3,4,5,6$. We use the method delineated in \cite{Banks:1989df}, and compare our results to recent literature. We propose a general structure for the genus-$h$ disk path integral, and comment on the large-$\ell$ behaviour, where $\ell$ is the proper size of the $S^1$ spatial boundary. In section \ref{wdwtop}, we discuss the failure of the higher-genus path integrals in solving the Wheeler-DeWitt equation, drawing an analogy to a simple quantum mechanical setting. In section \ref{outlook}, we compare our results to the more realistic case of timelike Liouville theory. Various details and elaborations are offered in the appendices.

\section{Two-dimensional quantum gravity}\label{The Continuum Approach at genus-zero}

In this section we discuss quantum gravity in two dimensions, endowed with a positive cosmological constant $\Lambda$. From a path integral perspective, one must integrate over smooth metrics, $g_{ij}$, on a closed genus-$h$ Riemann surface $\Sigma_h$. The most general local, diffeomorphism invariant, two-derivative action is given by
\begin{equation}\label{pure gravity action unfixed gauge}
S_{\text{grav}}[g_{ij};\Sigma_h]=-\frac{\vartheta}{4 \pi} \int_{\Sigma_h}\dd^2x\sqrt{g}R+\Lambda\int_{\Sigma_h}\dd^2x\sqrt{g}~,
\end{equation}
where $\vartheta$ and $\Lambda$ are real-valued positive parameters and $R$ is the Ricci scalar. By the Gauss-Bonnet theorem, the first term is a topological invariant proportional to the Euler characteristic $\chi_h$. The second term, which is the cosmological constant term, computes the area of the surface. We will also consider manifolds with boundaries.
\newline\newline
The theory (\ref{pure gravity action unfixed gauge}) is exceedingly simple, with no propagating gravitons to scatter, no semiclassical equations to solve, and no local ultraviolet divergences to tame. Nonetheless, the theory encodes non-trivial quantum fluctuations of the metric field, and admits a non-trivial gravitational path integral on a genus-$h$ surface
\begin{equation}\label{PI pure geometry}
\mathcal{Z}^{(h)}_{\text{grav}}[\Lambda] \equiv e^{\vartheta\chi_h}\int\frac{\mathcal{D}g_{ij}}{\text{vol} \, \mathcal{G}_{\text{diff}}}\,\exp\left({-\Lambda \int_{\Sigma_h}\dd^2x \sqrt{g}}\right)~.
\end{equation}
In addition, one can add matter fields and boundaries to the theory, adding further complexity. To retain the same degree of control as the pure gravitational theory, the matter theory is often chosen to be a minimal model (see \cite{Klebanov:1991qa,ginsparg1991large,anninos2020notes} for reviews). Here, we will focus exclusively on the pure case, as the questions we are interested in are already appear there and moreover generalize in a simple way when gravity is coupled to a minimal model.
\newline\newline
Before delving into the more detailed and complete structure of the quantum theory in section \ref{2dLiouville}, we will first discuss some properties from a more general `effective field theoretic' perspective.

\subsection{`Effective field theory' couplings}\label{infcouplings}

In an ordinary theory of gravity, in addition to the Einstein term, effective field theoretic considerations generally lead to an infinite set of additional diffeomorphism invariant higher-derivative terms. Naively, one might expect these to appear in the full quantum theory completing (\ref{pure gravity action unfixed gauge}). This would  indicate a significant number of observables/calculables for the two-dimensional theory. However, the simplicity of two-dimensional gravity undoes this structure, almost entirely.
\newline\newline
\textbf{Absence of bulk coupling constants.} In two spacetime dimensions there are no non-trivial tensor invariants and everything is built from the Ricci scalar, $R$. All effective field theoretic terms one might imagine adding can be removed by appropriate local field redefinitions of the metric. For instance, taking the field redefinition $g_{\mu\nu} \to g_{\mu\nu}+ \alpha f(R,\partial_\mu R,\ldots) g_{\mu\nu}$ with  a judiciously chosen $f$, allows us to remove any effective field theory couplings (except $\vartheta$ and $\Lambda$) in the effective field theoretic extension of (\ref{pure gravity action unfixed gauge}). Effectively, this means that (in the absence of boundaries) pure gravity carries only the meaningful local coupling constant, $\Lambda$, along with the topological coupling, $\vartheta$. All the other infinite gravitational couplings are redundant, in the sense of effective field theory. Thus, on a closed manifold, the gravitational path integral can only depend on $\Lambda$ and $\vartheta$. In fact upon fixing the genus, $h$, the $\mathcal{Z}^{(h)}_{\text{grav}}[\Lambda]$ can be used to define $\Lambda$ and $\vartheta$ in a field redefinition invariant way \cite{anninos2021semiclassical}.  
\newline\newline
\textbf{Boundary coupling constants.} In the presence of a boundary, new observables can potentially emerge, as well as new physical couplings. For instance, we can have a coupling, $\mu_B$, linked to the proper size of the boundary. As such, the gravitational path integral will also depend on some function of  the dimensionless ratio $\tfrac{\mu_B^2}{\Lambda}$. It is not unusual for a theory that is empty on a closed space to admit a richer structure on a manifold with a boundary; this can happen for topological quantum field theories \cite{Elitzur:1989nr}. From a gravitational perspective, we can organise the set of boundary couplings in terms of extrinsic and intrinsic quantities. For a one dimensional undecorated boundary, one only has  the intrinsic coupling $\mu_B$ associated to the proper length of the boundary. Extrinsic terms are more numerous since we can take the boundary integral over arbitrary powers of the extrinsic curvature, $K$, and its derivatives tangential to the boundary. Some of these boundary couplings may be linked to higher derivative bulk terms, as the latter will yield boundary terms upon integrating by parts. But others, such as $\mu_B$, are independent. One extrinsic boundary coupling is the Gibbons-Hawking-York \cite{york1972role,gibbons1977action} coupling, $\gamma$, controlling $K$. We can tune this coupling to ensure that when combined with the bulk $R$ term, one has a topological coupling $\vartheta$ on the manifold with a boundary. The boundary couplings $\mu_B$ and $\gamma$ are the only ones that are either relevant or marginal. All other couplings turn on irrelevant operators. 
\newline\newline
{This discussion has been a little general. To illustrate the main concepts for this type of problem, we provide a simple but concrete analysis of a shift symmetric scalar field on a manifold with a boundary in appendix \ref{EFTb}.}

\subsection{Matrix model perspective} 

%of a matrix potential admitting two-dimensional quantum gravity
The above discussion can also be understood through the relation of two-dimensional quantum gravity to an $N\times N$ matrix integral \cite{douglas1990strings, brezin1990exactly, gross1990nonperturbative}. Here, the continuum gravitational theory is realised by tuning the space of couplings of the matrix model at large $N$ (see \cite{ginsparg1991large,anninos2020notes} for reviews). As a simple example, we can take the matrix potential $V_\alpha(M) = N \text{Tr} \left( 
\tfrac{1}{2} M^2 + \alpha M^4 \right)$. Concurrently taking the large-$N$ limit along with $\alpha \to \alpha_{\text{cr}}=-\tfrac{1}{48}$ whilst keeping $\kappa^{-1} = N(\alpha-\alpha_{\text{cr}})^{-\frac{5}{4}}$ fixed is the desired double scaling limit (d.s.l.). The correspondence for partition functions reads
\begin{equation}\label{Zmatrix}
    \lim_{\text{d.s.l.}} \log\frac{\mathcal{M}_N(\alpha)}{\mathcal{M}_N(0)} \equiv \sum_{h=0}^\infty \mathcal{Z}^{(h)}_{\text{grav}}[\Lambda]~, \quad\quad\quad \mathcal{M}_N(\alpha) \equiv \int \dd M e^{-V_\alpha(M)}~.  
\end{equation}
The gravitational couplings are linked to $\alpha$ and $N$ in an explicit way. They group together into the parameter $\kappa^{-1} = e^{\vartheta}\Lambda^{\frac{5}{4}} $. In addition to the quartic coupling in $V_\alpha(M)$, one can turn on an infinite number of couplings $\lambda_n$, with $n \in \mathbb{Z}^+$, sourcing the single trace matrix monomials $\text{Tr} M^n$. Provided these couplings are small, they have no impact on the critical behaviour of the matrix model, and do not affect the invariant content of the matrix integral $\mathcal{M}_N(\alpha)$ in the double scaling limit. 
\newline\newline
We can also consider the perspective of the string equation for pure gravity \cite{douglas1990strings,brezin1990exactly,gross1990nonperturbative}, which encodes all genus corrections in the double scaling limit. For pure gravity, the string equation contains an infinite set of couplings, $\{z,t_1,t_2,\ldots\}$. The genus-zero string equation for pure gravity  is given by 
\begin{equation}\label{seqg}
 \mathcal{P}_0(z;t_n)^2 + \sum_{n>2} t_n \,  \mathcal{P}_0(z;t_n)^n - z = 0~,
%+ \frac{\kappa^2}{6} u''(z;t_n) 
\end{equation}
with the $t_n$ viewed as small couplings from the string equation perspective. To make contact with the gravitational picture, one picks the solution that at large-$z$ asymptotes to $\mathcal{P}_0(z;t_n)\approx -z^{\frac{1}{2}}$ encodes the structure of $\mathcal{Z}^{(0)}_{\text{grav}}[\Lambda]$. The parameter $z$ plays the role of $\Lambda$ (or, more precisely, $e^{\frac{4\vartheta}{5}}\Lambda$, up to some overall scheme dependent positive normalisation). The partition function  non-analytically depends on $z$ through the relation $\partial_z^2 \mathcal{Z}^{(0)}_{\text{grav}}[z] = \mathcal{P}_0(z;0)$. This yields $\mathcal{Z}^{(0)}_{\text{grav}}[\Lambda] = -\tfrac{4}{15}e^{2\vartheta}{\Lambda}^{\frac{5}{2}}$, up to terms analytic in $\Lambda$. What is observed in \cite{Moore:1991ir}, is that the remaining couplings, $t_n$, yield only analytic contributions to the partition function, rather than additional non-analyticities that have invariant meaning. These can be absorbed into an analytic redefinitions of  $z$.\footnote{A simple example suffices to see the general picture. Take the case where $t_3$ is switched on in (\ref{seqg}), with all other $t_k=0$. Expanding the exact solution that completes $-z^{\frac{1}{2}}$ in a small-$t_3$ expansion reveals terms that are analytic in $z$ and $t_3$. These are not universal. There are also terms that are non-analytic in $z$, that go as $z^\frac{7}{2}$ and higher. These can be grouped into the term $-z^{\frac{1}{2}}(1+\tfrac{5}{8}z \,t_3^2 + \ldots)$ which we can absorb into a redefinition of $z$ with the same underlying non-analyticity. We further note that the operator $t_2$  can be can be removed with a rescaling of $z$.}
\newline\newline
Having said that, once the normalisation of the leading genus-zero contribution is fixed, the pre-factors of the higher-genus contributions become meaningful and non-trivial theoretical data of the theory \cite{anninos2021semiclassical}.  %Beyond genus-zero, the string equation generalises to 
\newline\newline
\textbf{Hilbert space picture?} It would seem, at least in the absence of a boundary on a fixed topology, that we are studying a rather empty theory, with little physical content. On the other hand, even for this simplest of quantum gravities with $\Lambda>0$, a precise understanding of the quantum mechanical Hilbert space, if it exists at all, is lacking. This might have already been anticipated from the fact that the only microscopic description available for the theory stems from a matrix integral, built from $N^2$ c-numbers that have no underlying quantum mechanical operator algebra or Hilbert space structure. In contrast, from a continuum perspective nothing appears to preclude us from quantising the gravitational theory using canonical methods. In the following sections we will discuss these issues from the more explicit perspective of Liouville theory.

\section{Pure gravity as a Liouville CFT}\label{2dLiouville}

An efficient and more complete way to study the theory (\ref{pure gravity action unfixed gauge}) at the quantum level is to select the gauge 
\begin{equation}\label{conformal gauge}
g_{ij}(x)=e^{\varphi(x)}\Tilde{g}_{ij}(x)~.
\end{equation}
Here, $\Tilde{g}_{ij}(x)$ is a fixed reference metric, often referred to as the fiducial metric, and $\varphi(x)$ is a local Weyl factor that remains unfixed. The parametrization (\ref{conformal gauge}) does not entirely fix the gauge as it leaves a particular subgroup of diffeomorphisms unfixed. Moreover, the transformation
\begin{equation}\label{residual Weyl diffeo}
\Tilde{g}_{ij}(x) \rightarrow e^{\Tilde{\sigma}(x)}\Tilde{g}_{ij}(x), \quad\quad \varphi(x) \rightarrow \varphi(x)-\Tilde{\sigma}(x)
\end{equation}
for general $\Tilde{\sigma}(x)$ is a redundancy of (\ref{conformal gauge}).  As such, the resulting theory governing $\varphi(x)$ and the accompanying Fadeev-Popov ghost fields should be invariant under the more general Weyl transformation (\ref{residual Weyl diffeo}) \cite{distler1989conformal,david1988conformal}. In the gauge (\ref{conformal gauge}), the ghost theory is described by a conformal field theory of $\mathfrak{bc}$-ghosts in a fixed background $\Tilde{g}_{ij}(x)$ whose central charge is $c_{\text{gh}}=-26$.

\subsection{Basic ingredients}\label{basic}

If one assumes that the $\varphi$-sector admits a local description, one can postulate an action of the following type \cite{david1988conformal, distler1989conformal}
\begin{equation}\label{Liouville action}
S_\text{L} =\frac{1}{4\pi}\int_{\Sigma_h}\dd^2x\sqrt{\Tilde{g}}\left( \Tilde{g}^{ij}\partial_i\varphi\partial_j\varphi+Q\Tilde{R}\varphi+4\pi\Lambda e^{2b\varphi}\right).
\end{equation}
The  Liouville action has been studied extensively \cite{nakayama2004liouville,seiberg1990notes,teschner2001liouville,zamolodchikov2007lectures,fateev2000boundary}.\footnote{We will use the language of fields throughout to guide the discussion, but at a finite value of $b$ the Liouville theory is best understood as an inherently quantum theory defined either through a rigorously treated path integral or conformal bootstrap methods.}
In order for the kinetic term to have a standard normalisation, we have allowed for a constant rescaling of $\varphi$ by $2b$. When $Q=b+{b^{-1}}$ the Liouville action describes a two-dimensional conformal field theory \cite{teschner2001liouville,Dorn:1994xn,zamolodchikov1996conformal}. It has  a continuous spectrum of primary operators and the absence of a normalizable $PSL(2,\mathbb{C})$ invariant vacuum state. Nevertheless, it admits a consistent quantization and many of its properties are known explicitly. For instance, its central charge is given by $c_{\text{L}} = 1+6Q^2$, and the theory contains spinless primary operators, $\mathcal{O}_{\alpha}=e^{2\alpha \varphi}$, of conformal dimension
\begin{equation}\label{con dim primaries}
\Delta_\alpha = \bar{\Delta}_\alpha = \alpha(Q-\alpha)~. % \quad\quad  \alpha(Q-\alpha)~.
\end{equation}
In turn, these conformal dimensions can be used to fix $b$ in (\ref{Liouville action}), since conformal invariance requires $\mathcal{O}_{b}$ to have conformal dimensions $(\Delta_b,\bar{\Delta}_b)=(1,1)$. One thus finds
\begin{equation}\label{b and Q relation}
b=\frac{Q}{2} \pm \sqrt{\frac{Q^2}{4}-1}~.
\end{equation}
Requiring $c_{\text{L}}+c_{\text{gh}}=0$ fixes $Q=\pm \frac{5}{\sqrt{6}}$. Given the discrete symmetry $Q\rightarrow -Q, \; b\rightarrow-b$ and $\varphi\rightarrow-\varphi$, we can pick $Q>0$ without loss of generality. One finds two solutions, $b=b_\pm$ with $b_+=\sqrt{\frac{3}{2}}$ and $b_-=\sqrt{\frac{2}{3}}$. To make contact with the matrix model results, one is required to pick the negative root in (\ref{b and Q relation}). We will consider the negative root in what follows, such that  $b=\sqrt{\frac{2}{3}}$ and $Q=\tfrac{5}{\sqrt{6}}$.
\newline\newline
A natural collection of observables is given by the set of diffeomorphism invariant functionals of $g_{ij}$. In the Weyl gauge, this is given by the set of conformal primaries built from $\varphi$ with conformal dimensions $(\Delta,\bar{\Delta})=(1,1)$. A particularly important example is the area operator, equally the integrated $\mathcal{O}_{b}$ operator,
\begin{equation}\label{Area operator}
\mathcal{A}_h=\int_{\Sigma_h}\dd^2x \sqrt{g}=\int_{\Sigma_h}\dd^2x\sqrt{\Tilde{g}}e^{2b\varphi}~.
\end{equation}
The non-analytic dependence on $\Lambda$ of (\ref{PI pure geometry}) is given by \cite{distler1989conformal, david1988conformal} 
\begin{equation}\label{PI Lambda dep}
\mathcal{Z}^{(h)}_{\text{grav}}[\Lambda] = \mathcal{N}_h \,e^{\vartheta \chi_h}\left(\frac{\Lambda}{\Lambda_{\text{u.v.}}}\right)^{\frac{Q\chi_h}{2b}}~,  \quad\quad h\neq1~,%\,\Gamma\left(-\frac{Q\chi_h}{2b}\right) \,
\end{equation}
where in the case of closed manifolds $\chi_h=2-2h$.  When $h=1$ the resulting expression is logarithmic in $\Lambda$ whilst independent of $\vartheta$ \cite{Bershadsky:1990xb}. We note that one must introduce an ultraviolet scale $\Lambda_{\text{u.v.}} \equiv \ell_{\text{u.v.}}^{-2}$ to define the gravitational path integral, and constant changes in its value can be absorbed into redefinitions of $\vartheta$ and/or $\Lambda$. For pure two-dimensional gravity we have $\mathcal{Z}^{(h)}_{\text{grav}}[\Lambda] = \mathcal{N}_h (e^{\vartheta}\Lambda^{\frac{5}{4}})^{\chi_h}$ for $h\neq 1$, while for $h=1$ we have $\mathcal{Z}^{(1)}_{\text{grav}}[\Lambda] = -\tfrac{1}{48} \log \Lambda$ which is independent of $\vartheta$ \cite{Bershadsky:1990xb}. We now comment on the pre-factor $\mathcal{N}_h$.

\subsection{Comment on the sign of $\mathcal{Z}^{(h)}_{\text{grav}}[\Lambda]$}\label{signZ}

Matching the pure gravity result for $\mathcal{Z}^{(0)}_{\text{grav}}[\Lambda]$ to the matrix model expression, we must further take $\mathcal{N}_0$ in (\ref{PI Lambda dep}) (whose precise value depends on the details of the model) to be negative. {Take again, as a concrete example, the quartic matrix potential $V_\alpha(M) = N\text{Tr}\left(\tfrac{1}{2}M^2+\alpha M^4\right)$. This is analysed, for example, in section 3.2 of \cite{anninos2020notes}. One  finds that near the critical coupling, $\alpha_{\text{cr}}=-\tfrac{1}{48}$, the non-analytic contribution to the planar approximation of the free energy is $\partial_\alpha^2 (-N^2 \mathcal{F}^{(0)}_N(\alpha)) \approx -9216 N^2 \sqrt{3(\alpha -\alpha_{\text{cr}})}$, where $\mathcal{F}_{N}(\alpha)\equiv-\log\frac{\mathcal{M}_N(\alpha)}{\mathcal{M}_N(0)}$. The leading non-planar correction is $-\mathcal{F}^{(1)}(\alpha) \approx -\tfrac{1}{24} \log{(\alpha -\alpha_{\text{cr}})}$, and this matches the torus partition function $\mathcal{Z}^{(1)}_{\text{grav}}[\Lambda]$ computed in \cite{Bershadsky:1990xb}. Again the match includes the sign, and for the torus partition function we have a trace interpretation of the continuum calculation. (The mismatch in the coefficient of the logarithmic terms at genus one, $-\tfrac{1}{24}$ versus $-\tfrac{1}{48}$ in \cite{Bershadsky:1990xb}, is due to the choice of an even matrix potential that effectively doubles the system \cite{Bachas:1990ua}.) So the matrix model correlates the signs of $\mathcal{Z}^{(0)}_{\text{grav}}[\Lambda]$ and $\mathcal{Z}^{(1)}_{\text{grav}}[\Lambda]$. The same sign follows directly from an analysis of the string equation for pure gravity (for example, see (2.28) of \cite{DiFrancesco:1993cyw}).  When translating back to the gravitational picture, the cosmological constant $\Lambda$ goes as $(\alpha-\alpha_{\text{cr}})$, and $\vartheta$ goes as $\log N$. Beyond the non-planar contributions, one also has a non-perturbative contribution due to an eigenvalue displaced from the dense distribution to a local maximum of its effective potential. This contributes a pure imaginary term (as analyzed in \cite{Hanada:2004im}) to the partition function.   
\newline\newline
At this stage, one might be still inclined to argue that the sign of $\mathcal{Z}^{(0)}_{\text{grav}}[\Lambda]$ is  ambiguous, or a matter of convention. From the matrix model viewpoint, however, the non-analyticities in $\alpha$ are a consequence of the asymptotic number of graphs \cite{brezin1990exactly,Eguchi:1982fe,Zamolodchikov:1982vx,anninos2021matrix}, $N_{\text{g}}$, on $S^2$ with $n$ vertices. These grow as $N_{\text{g}} \sim \exp (c_{\text{g}}  n) n^{-\frac{7}{2}}$ where $c_{\text{g}}$ is a model dependent coefficient. So if we wish to give a statistical/counting interpretation to $\mathcal{Z}^{(0)}_{\text{grav}}[\Lambda]$, as envisioned by Gibbons and Hawking \cite{gibbons1978path}, it again seems that we should select $\mathcal{Z}^{(0)}_{\text{grav}}[\Lambda]$ to be negative. Finally, we can also compute $\mathcal{Z}^{(0)}_{\text{grav}}[\Lambda]$ using the DOZZ formula \cite{dorn1994two,Zamolodchikov:1995aa} for the structure constant, $C_{bbb}$, of three area operators $\mathcal{O}_b$. We discuss this approach in section \ref{normLZ}, where we once again find a negative $\mathcal{Z}^{(0)}_{\text{grav}}[\Lambda]$  for pure two-dimensional quantum gravity with $\Lambda>0$.
\newline\newline
If we consider two-dimensional $\Lambda>0$ quantum gravity coupled to a general minimal model, the pre-factor for the gravitational path integral generalises (see, for instance, (8.18) of \cite{anninos2020notes}). It can be positive or negative depending on the central charge of the minimal model. $\mathcal{Z}^{(0)}_{\text{grav}}[\Lambda]$ appears to be negative for unitary minimal model matter fields (see \cite{Spodyneiko:2014lla,Duits:2024bej,Hayford:2024whf} for a recent analysis of the Ising model coupled to quantum gravity). Another continuum method is to compute the fixed area partition function of the gravitational theory \cite{Zamolodchikov:1982vx,Muhlmann:2021clm}. Provided we take this to be positive, $\mathcal{N}_0$ is again found to be negative (see for example equation (8.18) of \cite{anninos2020notes}). A negative $\mathcal{Z}^{(0)}_{\text{grav}}[\Lambda]$ indicates no pathology --- it is a regularised quantity contributing to the logarithm of the matrix integral (\ref{Zmatrix}). Indeed, at genus zero there can also be divergent contributions from the fixed area partition function that yield analytic terms in $\Lambda$. Though positive, these are non-universal.\footnote{It would also be interesting to study the case of two-dimensional supergravity with $\Lambda>0$, which can exhibit novel features due to the appearance of Ramond-Ramond sectors \cite{Klebanov:2003wg}. Here, the matrix model is a rectangualar $(N+\Gamma)\times N$ matrix model yielding a slightly different combinatorial problem that also depends on $\Gamma$ which is interpreted in terms of the Ramond-Ramond flux and yields additional decorations on the $S^2$.}
\newline\newline
Finally, it worth noting that the sphere path integral of four-dimensional $\Lambda>0$ Einstein gravity, at one-loop, is also negative \cite{polchinski1989phase}. For more exotic theories of four-dimensional gravity with higher-spins, the phase can be different, and even vanish \cite{Giombi:2026sqa,Anninos:2026hia}. Beyond the sphere, one can find four-dimensional gravitational instantons whose phase is different from that of the sphere \cite{Anninos:2025ltd}, vaguely reminiscent of the imaginary terms uncovered in \cite{Hanada:2004im} contributing non-perturbatively to the two-dimensional partition function. 

\subsection{Manifold with a boundary}

Mindful of our subsequent discussion, let us also consider Liouville theory \cite{fateev2000boundary, nakayama2004liouville} on a manifold with a  single boundary. A conformally invariant boundary condition that supplements the bulk term (\ref{Liouville action})
can be introduced though the following boundary interaction
\begin{equation}\label{Boundary Liouville theory}
S_{\text{L,bdy}} = \frac{1}{2 \pi} \int_{\partial \Sigma_h} \dd\xi \sqrt{\tilde{h}}\left({Q \tilde{K} \varphi}+2\pi\mu_Be^{b\varphi}\right)~,
\end{equation}
where the integration in $\xi$ is along the boundary with induced metric $\tilde{h}$, while $\tilde{K}$ is the trace of the extrinsic curvature of the boundary embedded in the background geometry $\Tilde{g}$. Thus, there is a one-parameter family of conformally invariant boundary conditions characterized by different values of $\mu_B$ \cite{fateev2000boundary} referred to the FZZT boundary condition. An alternative boundary condition was given in \cite{Zamolodchikov:2001ah}, labeled by two positive integers $(m,n)$ that enforce $e^{b\varphi} \to \infty$ at the boundary. For the case of pure two-dimensional gravity we further have $(m,n)=(1,1)$. In \cite{Seiberg:2004at} it is shown how the two types of boundary conditions are related to each other.
\newline\newline
For a manifold $\Sigma_{h,\text{b}}$ of genus $h$ with $\text{b}$ boundaries, one has an Euler characteristic $\chi_{h,\text{b}} = 2-2h-\text{b}$. We are interested in the case $\text{b}=1$, for which the corresponding partition function takes the general form
\begin{equation}\label{PI Lambda dep bndry}
\mathcal{Z}^{({h,1})}_{\text{grav}}[\Lambda;\mu_B] =  e^{\vartheta \chi_{h,1}}f_h\left( \frac{\mu_B^2}{\Lambda} \right) \left(\frac{\Lambda}{\Lambda_{\text{u.v.}}}\right)^{\frac{Q\chi_{h,1}}{2b}}~.
\end{equation}
Contrary to the pure bulk situation where cosmological constant enters as the only dimensionful quantity, the observables in the boundary case can further depend on the dimensionless ratio $\frac{\Lambda}{\mu_B^2}$. The basic boundary operator primaries are again exponentials in $\varphi$ boundary fields $\mathcal{B}_\beta \equiv e^{\beta \varphi}.$ The boundary conformal weight of $\mathcal{B}_\beta$ is given by \cite{fateev2000boundary}
\begin{equation}\label{conformal weight boundary operator}
\Delta_\beta =  \bar{\Delta}_\beta = \beta(Q-\beta)~. %, \quad\quad \beta(Q-\beta)~.
\end{equation}
To avoid confusion, we use the parameter $\alpha$ for the bulk exponential and parameter $\beta$ in relation with the boundary operators. Another important operator is the boundary length operator, the integrated version of $\mathcal{B}_{\beta}$,
\begin{equation}\label{Boundary operator}
\mathcal{B}_h = \int_{\partial \Sigma_h}\dd\xi \sqrt{\tilde{h}} e^{b\varphi}~.
\end{equation}
The Euclidean quantity, $\mathcal{Z}^{({h,1})}_{\text{grav}}[\Lambda;\mu_B]$, may stem from a variety of Lorentzian perspectives. One is related to the Hartle-Hawking preparation of wavefunctions on a closed spatial slice in terms of a no boundary path integral \cite{hartle1983wave}. Here, the boundary is viewed as the closed spatial Cauchy surface of the Lorentzian theory. Another interpretation is the Euclidean continuation of a Lorentzian theory with a timelike boundary. In this case, the Lorentzian theory resides on a spatial half-line (or an interval if we have two boundaries). To obtain a function of the proper length of the boundary, $\ell$, we can perform an inverse Laplace transform for $\mu_B$.

\section{BRST cohomology}\label{LZsec}

In this section, we review the basic building blocks of the gravitational Hilbert space, $\mathcal{H}_{\text{phys}}$, from the perspective of the relative BRST cohomology  \cite{lian1991new,lian1991semi,lian1992semi}. This can be viewed as a canonical construction of the quantum mechanical Hilbert space of the gravitational theory on a spatial circle, having implemented the conformal gauge. We discuss subtleties in defining an inner product for such a Hilbert space.

\subsection{Description of the cohomology}

Upon imposing the Weyl gauge, (\ref{conformal gauge}), we must incorporate the $\mathfrak{bc}$-ghost theory into the underlying description of the physical Hilbert space, $\mathcal{H}_{\text{phys}}$. Physical states are elements of the BRST cohomology in the combined Liouville and ghost Hilbert space  $\mathcal{H}_{\text{L}} \otimes \mathcal{H}_{\text{ghost}}$. When viewed as an ordinary conformal field theory, states in the Liouville Hilbert space map to operators of the type $\mathcal{O}_\alpha = e^{2\alpha \varphi}$ with $\alpha = \tfrac{Q}{2}+ i P$ with $P \in \mathbb{R}$ and $\Delta_\alpha = \tfrac{Q^2}{4} + P^2$.  These operators create $\delta$-function normaliseable eigenstates of the Hamiltonian. However, from the perspective of the gravitational theory in conformal gauge, one is left with a set of residual redundancies that further enforce physical states (and operators) to be invariant under the conformal symmetries. In particular, the combined conformal weight of the ghost and Liouville operators must vanish. This  requires that $P \notin \mathbb{R}$, forcing us to extend beyond the ordinary collection of Liouville operators.
\newline\newline
Thus, the continuum of operators that are physical from the perspective of Liouville theory, viewed as a unitary quantum field theory, need not directly correspond to the appropriate physical picture in the two-dimensional quantum gravity theory.
Moreover, Lian and Zuckerman showed \cite{lian1991new, lian1992semi}  that the BRST cohomology consists of an infinite number of operators containing non-trivial ghost number, denoted by $n_{\text{gh}}$ and $\tilde{n}_{\text{gh}}$. Taken at face value, our essentially `empty' theory of quantum gravity now suffers from an embarrassment of riches --- a potentially infinite dimensional gravitational Hilbert space. The structure of the Lian-Zuckerman operators for pure two-dimensional quantum gravity is given by
\begin{equation}\label{Lian-Zuckerman operators}
\mathcal{R}^{\text{LZ}}_{\pm}(t) \equiv  \mathcal{O}^{\text{LZ}}_{\pm} \otimes \mathcal{\tilde{O}}^{\text{LZ}}_{\pm}\otimes e^{2 \alpha_t \varphi}~,
\end{equation}
where $t\in\mathbb{Z}$ and $\pm$ labels the particular Lian-Zuckerman operator, $\mathcal{O}^{\text{LZ}}_{\pm}$, which is built from the ghost and Liouville fields. To ensure gauge invariance under the residual gauge freedom,   the holomorphic  scaling dimensions of the operators (\ref{Lian-Zuckerman operators})  must satisfy
\begin{equation}\label{Lian-Zuckerman grav dressing condition}
\Delta^{\text{LZ}}_{\pm}(t)+\alpha_t(Q-\alpha_t)=0~,
\end{equation}
and similarly for the anti-holomorphic sector. The Lian-Zuckerman weights are given by (see for example \cite{lepowsky2013vertex, lian1991new})
\begin{equation}\label{Lian-Zuckerman weights}
\Delta^{\text{LZ}}_{+}(t) =  6 t^2 + 5 t~, \quad\quad \Delta^{\text{LZ}}_{-}(t) = 6 t^2-t-1~.
%\footnote{The anti-holomorphic conformal dimension has to be equal to the holomorphic conformal dimension.}
\end{equation}
The argument $t \in \mathbb{Z}$ is related to the ghost number through the embedding diagram \cite{lian1991new}, whereas the subscript $\pm$ indicates whether the ghost number of the operator  is even $(+)$ or odd $(-)$. Low lying examples of Lian-Zuckerman operators (up to BRST exact terms) are given by \cite{imbimbo1992construction,kanno1994brst, mukhi1992extra, govindarajan1992states, alekseev2010ring, anninos2021matrix}
\begin{eqnarray}\label{identity Lian-Zuckerman state} \nonumber
n_{\text{gh}} = \tilde{n}_{\text{gh}} = 1: \quad\quad & &\mathcal{R}^{\text{LZ}}_{-}(0) =\mathfrak{c}(z)\tilde{\mathfrak{c}}(\bar{z}) e^{2b\varphi}~,\\ \nonumber
n_{\text{gh}} = \tilde{n}_{\text{gh}} =  0: \quad\quad & &\mathcal{R}^{\text{LZ}}_{+,0}(0) = \mathbb{I}~, \\ \nonumber
n_{\text{gh}} =  \tilde{n}_{\text{gh}} = 2:  \quad\quad & &\mathcal{R}^{\text{LZ}}_{+,2}(0)= \mathfrak{c}(z)\partial_z^2 \mathfrak{c}(z) \tilde{\mathfrak{c}}(z) \partial_{\bar{z}}^2 \tilde{\mathfrak{c}}(\bar{z}) e^{2Q\varphi}~,\\ \nonumber
n_{\text{gh}} =  \tilde{n}_{\text{gh}} = 0:  \quad\quad & &\mathcal{R}^{\text{LZ}}_{+,0}(-1)=\left(\mathfrak{b}(z)\mathfrak{c}(z)-\frac{\partial\varphi(z,\bar{z})}{{b}}\right) \left(\tilde{\mathfrak{b}}(\bar{z})\tilde{\mathfrak{c}}(\bar{z})-\frac{\bar{\partial}\varphi(z,\bar{z})}{b}\right) e^{-b\varphi(z,\bar{z})}~, \\ 
n_{\text{gh}} =  \tilde{n}_{\text{gh}} = 2:  \quad\quad & &\mathcal{R}^{\text{LZ}}_{+,2}(-1)= \mathfrak{c}(z)\partial_z^3 \mathfrak{c}(z) \tilde{\mathfrak{c}}(z) \partial_{\bar{z}}^3 \tilde{\mathfrak{c}}(\bar{z}) e^{\frac{4}{b}\varphi}~,
\end{eqnarray}
with additional constructions found in \cite{imbimbo1992construction,kanno1994brst,govindarajan1992states}. We recall that $Q=\frac{5}{\sqrt{6}}$ and $b=\sqrt{\frac{2}{3}}$ for pure two-dimensional gravity.
%More generally, the ghost number of the Lian-Zuckerman operator $\mathcal{R}^{\text{LZ}}_{\pm}(t)$ is given by \cite{lian1991new}. 
The corresponding Lian-Zuckerman states are obtained by acting with the Lian-Zuckerman operators on the tensor product of the Liouville vacuum $\ket{0_{\text{L}}}$ and the $PSL(2,\mathbb{C})$ invariant ghost state $\ket{0_{\text{gh}}}$ which we define to carry $n_{\text{gh}} = \tilde{n}_{\text{gh}} =-1$.

\subsection{CFT inner-product of Lian-Zuckerman states}\label{normLZ}

One  would also like to define an inner product between two elements in the cohomology and find their norm. This is subtle for both the Liouville and ghost sectors.
\newline\newline
{\textbf{Liouville norm.}} The Liouville vertex operators in (\ref{Lian-Zuckerman operators}) have imaginary $P$, and therefore do not correspond to the standard $\delta$-function normalisable states in the quantum field theoretic Liouville Hilbert space. To define their inner product, we consider their two-point function on $S^2$, with the two operators inserted at antipodal points. The two-point function can be obtained from the analytically extended expression of Dorn and Otto \cite{dorn1994two} and Zamolodchikov and Zamolodchikov \cite{zamolodchikov1996conformal} (DOZZ) for the general structure constant $C_{\alpha_1 \alpha \alpha_2}$ of the Liouville conformel field theory. Setting one of the vertex operators to become the identity yields \cite{teschner2001liouville}
\begin{equation}\label{L2pt}
\lim_{\alpha \to 0} C_{\alpha_1 \alpha \alpha_2} =  2 \pi  \delta(\alpha_1+\alpha_2-Q)+ S(\alpha_1)\delta(\alpha_1-\alpha_2)~,
\end{equation} 
where $S(\alpha)$ is the reflection amplitude given by
\begin{equation}\label{Reflection amplitude}
    S(\alpha) \equiv \frac{\left(\pi \Lambda\gamma(b^2)\right)^{\frac{Q-2\alpha}{b}}}{b^2}\frac{\gamma(2b\alpha-b^2)}{\gamma(2-2b^{-1}\alpha+b^{-2})}~,
\end{equation}
and $\gamma(x)\equiv\frac{\Gamma(x)}{\Gamma(1-x)}$.
The $\delta$-function is a reflection of the non-compact nature of Liouville theory, and has been interpreted as the volume of the dilation subgroup of $PSL(2,\mathbb{C})$ \cite{seiberg1990notes}.
\newline\newline
{\textbf{Ghost norm.}} The inner product on the $\mathfrak{bc}-$ghost Hilbert space can be obtained from a similar treatment to that of the critical string   \cite{polchinski2005string}. Again, we consider it to be the two-point function for the ghost operator and its conjugate inserted at antipodal points of $S^2$. Here, one runs into trouble due to the presence of ghost zero modes. For instance, the norm of $\ket{0_{\text{gh}}}$ as computed by the $S^2$ path integral vanishes. By the Riemann-Roch theorem, this will be the case for any insertion that does not satisfy $n_{\text{gh}} = \tilde{n}_{\text{gh}} = 3$. From a more geometrical perspective, the vanishing stems from the necessity to divide by the volume of $PSL(2,\mathbb{C})$ or a non-compact subgroup thereof. Whether or not this is the case, the necessity to have a net ghost number of three is very restrictive.
\newline\newline
So in trying to define an inner product, we find ourselves in a $\infty_{\text{L}} \times 0_{\text{gh}}$ type situation. Not insurmountable, but far from ideal. 
To be concrete, let us compute the norm of the a Lian-Zuckerman state. We start with
\begin{equation}\label{norm0}
    \begin{split}
 \braket{0  _\text{L}|\, \braket{0{_\text{gh}}| \left(\mathcal{R}^{\text{LZ}}_{-}(0) \right)^\dag \mathcal{R}^{\text{LZ}}_{-}(0) |0_{\text{gh}}}  |0_{\text{L}}}
   &= \lim_{\alpha \to 0} C_{b  \alpha  b} \times  \bra{0_{\text{gh}}}\Tilde{\mathfrak{c}}_{-1}\mathfrak{c}_{-1}\mathfrak{c}_{1}\Tilde{\mathfrak{c}}_{1}\ket{0_{\text{gh}}} \\
    &=
    -\sqrt{\Lambda}\times\frac{12 \pi^{{5}/{4}}}{\Gamma(\frac{1}{6})^{{3}/{2}}} \times \delta(0) \times 0~.
    %\textcolor{blue}{S(b)} \times \delta(0) \times  0~.
    \end{split}
\end{equation}
The question is whether this $\infty_{\text{L}} \times 0_{\text{gh}}$ can be made sense of. One way to do so is to interpret the $0_{\text{gh}}$ as arising from dividing by the volume of the dilatation subgroup of $PSL(2,\mathbb{C})$ that remains unfixed after selecting two points on $S^2$.  Following \cite{seiberg1990notes}, the $\delta(0)$ infinity goes like the volume of the dilatation group. So it does not seem entirely unreasonable to regularise the above norm to some finite value. Finally, we note that the minus sign in the pre-factor of (\ref{norm0}) is not in conflict with the unitarity of spacelike Liouville theory, because we are evaluating the two-point function of an `unphysical' vertex operator with imaginary $P$.\footnote{A similar analysis yields a minus sign for the norm of the gravitational dressed primaries of the Ising model coupled two $\Lambda>0$ gravity. Crucially, they all have the same sign, so we can still view the theory as having a sign-definite inner product.}
\newline\newline
We also note that the subsequent Lian-Zuckerman states have vanishing norm in the following sense
\begin{equation} \label{R0}
\braket{0{_\text{L}}|\braket{0_{\text{gh}}|  \left(\mathcal{R}^{\text{LZ}}_{+,n}(0)  \right)^\dag \mathcal{R}^{\text{LZ}}_{+,n}(0) |0_{\text{gh}}}  |0_{\text{L}}}  = 0~,  \quad\quad n=0,2~.
% \braket{0 {_\text{L}}|\braket{0_{\text{gh}}| \left(\mathcal{R}^{\text{LZ}}_{+,2}(-1) \right)^\dag \mathcal{R}^{\text{LZ}}_{+,2}(-1) |0_{\text{gh}}}  |0_{\text{L}}}  &=& 0~.
\end{equation}
In fact, a simple application of the Riemann-Roch theorem will yield a vanishing norm for all subsequent Lian-Zuckerman states. Nonetheless, there are pairs of Lian-Zuckerman operators involving Liouville vertex operators of the type $\mathcal{O}_\alpha$ and $\mathcal{O}_{Q-\alpha}$ that may have a non-vanishing overlap. This can be checked explicitly for the pair $\mathcal{R}^{\text{LZ}}_{+,0}(-1)$ and $\mathcal{R}^{\text{LZ}}_{+,2}(-1)$. As such, the higher ghost number Lian-Zuckerman states need not be null states that can be removed. A related discussion can be found in \cite{imbimbo1992construction} (see their eqn. (6.9)).
\newline\newline
{\textbf{Gravitational path integral perspective.}} We can also consider the norm (\ref{norm0}) from the perspective of the gravitational path integral on $S^2$. Consider the two-point function obtained by taking two $\Lambda$ derivatives of the partition function $\mathcal{Z}_{\text{grav}}^{(0)}[\Lambda]$. This brings down two integrated cosmological operators, $\mathcal{O}_{b}$, into the path integral. One can use the residual $PSL(2,\mathbb{C})$ gauge redundancy to fix the two positions. This fixes two of the three non-compact directions of $PSL(2,\mathbb{C})$, leaving the non-compact dilatations unfixed. Recall, further, that the residual $PSL(2,\mathbb{C})$ is associated to the six ghost zero modes in the $\mathfrak{bc}$-ghost theory. Thus, from the $\mathfrak{b}\mathfrak{c}$-ghost point of view, we can replace the two integrals by dressing the vertex operators with a $\mathfrak{c}$ and $\tilde{\mathfrak{c}}$ ghost field. The remaining $\delta$-function divergence from the Liouville two point function can be canceled by the volume of the dilatation group in the denominator. From a gauge fixing perspective, we can instead further fix the dilatation group by inserting an additional $\mathfrak{c}$ ghost mode into the correlators. This restricted Liouville path integral will no longer yield two point functions with a $\delta(0)$. The decorated ghost norm is now finite, namely $\braket{0_{\text{gh}}|\tilde{\mathfrak{c}}_{-1}\mathfrak{c}_{-1}\tilde{\mathfrak{c}}_0 \mathfrak{c}_0 \tilde{\mathfrak{c}}_1\mathfrak{c}_{1}|0_{\text{gh}}}\equiv 1$, as for the critical string \cite{polchinski2005string}.
\newline\newline
Based on the above reasoning, we view the two point function (\ref{norm0}) as coming from two $-\Lambda$ derivatives of $\mathcal{Z}_{\text{grav}}^{(0)}[\Lambda]$. Employing the sphere path integral (\ref{PI Lambda dep}), we obtain
\begin{equation}\label{2ptRR}
\braket{0{_\text{L}}|\braket{0_{\text{gh}}|  \left(\mathcal{R}^{\text{LZ}}_{-}(0)  \right)^\dag \mathcal{R}^{\text{LZ}}_{-}(0) |0_{\text{gh}}}  |0_{\text{L}}}  = c_0 \sqrt{\Lambda}~,
\end{equation}
up to an overall normalisation dependent pre-factor, $c_0$, which again can be fixed through the DOZZ structure constant $C_{bbb}$. Using the conventions of \cite{teschner2001liouville}, for $b=\sqrt{\tfrac{2}{3}}$ we have
\begin{equation}
C_{bbb} = \frac{\sqrt{6}\pi^{1/4}}{\Gamma \left(\frac{1}{6}\right)^{\frac{3}{2}}}\frac{1}{\sqrt{\Lambda }}~.
%C_{bbb} = \frac{256\ 2^{\frac{5}{6}} \sqrt{3} \sqrt[4]{\pi }}{e^{\frac{25}{6}}  \Gamma \left(\frac{1}{6}\right)^{\frac{3}{2}}}\frac{1}{\sqrt{\Lambda }}~.
\end{equation}
As such discussed, the pre-factor on the right hand side of (\ref{2ptRR}) is not guaranteed to be positive. And indeed, using $C_{bbb}$ to compute $\mathcal{Z}_{\text{grav}}^{(0)}[\Lambda]$ by integrating thrice with respect to $-\Lambda$ yields a negative $c_0$.
\newline\newline
As was earlier mentioned, yet another method to compute $\mathcal{Z}^{(0)}_{\text{grav}}[\Lambda]$ is to first fix the area of $S^2$, and then integrate over the areas \cite{Zamolodchikov:1982vx,Muhlmann:2021clm}. Here, the negativity of the pre-factor in (\ref{2ptRR}) is tied to the positivity of the fixed area partition function.  To go back to the complete partition function, one has to regulate the area integral, by analytic continuation or cutting it off somehow, since it is otherwise divergent (see, for example, section 8.3 of section \cite{anninos2020notes}). The divergent terms are analytic in $\Lambda$. Finally, as discussed in section \ref{basic}, a direct matrix model computation \cite{brezin1990exactly} also yields a negative value for $\partial_\Lambda^{2}\mathcal{Z}^{(0)}_{\text{grav}}[\Lambda]$. Here,  the positivity properties stem from the fact that the Laplace transform of $\mathcal{Z}^{(0)}_{\text{grav}}[\Lambda]$ yields a counting problem for the number of graphs at a given topology, which we take to be positive.
%Perhaps we can make sense of the negative sign in $\mathcal{Z}^{(0)}_{\text{grav}}[\Lambda]$ as a small contribution to an overall posotive quantity sensitive to the divergent terms that have been removed.
\newline\newline
The Lian-Zuckerman states can also affect the expressions at higher genus. For instance, the genus-one partition function is sensitive to the Lian-Zuckerman spectrum \cite{Kutasov:1990sv,anninos2021matrix}, and at higher genus they may propagate internally in a type of gravitational sewing formulae. This may tie to the fact that once the genus-zero normalisation is fixed, the pre-factors of the higher-genus free energies are meaningful data of the matrix model.

\subsection{Timelike boundary, briefly}

An alternative Hilbert space for the theory can also be considered. This arises from quantising on a spatial slice with boundaries, rather than a spatial $S^1$. 
\newline\newline
{\textbf{Single boundary.}} From this point of view, we view the Euclidean disk path integral as a type of thermal trace, with the boundary $S^1$ being the thermal cycle. For pure gravity with fixed boundary length, $\ell$, the disk path integral with {no insertions} is given by \cite{fateev2000boundary,Moore:1991ir}
\begin{equation}\label{ZdiskU}
\mathcal{Z}_{D^2}[\tilde{\ell}] =  e^{\vartheta}  \left(\frac{\Lambda}{\Lambda_{\text{u.v.}}} \right)^{\frac{5}{4}}  {\left( \tilde{\ell}^{-\frac{5}{2}} +2 \tilde{\ell}^{-\frac{3}{2}}\right)}e^{-2\tilde{\ell} } ~, \quad\quad \tilde{\ell} \equiv \sqrt{\Lambda} \ell~. %\frac{Q}{2b}
\end{equation}
A simple derivation of the above follows by differentiating with respect to $\Lambda$. One then finds the disk path integral with an area operator inserted, (\ref{Area operator}), which reads \cite{fateev2000boundary,Moore:1991ir}
\begin{equation}
\mathcal{Z}^{\mathcal{(A)}}_{D^2}[\tilde{\ell}] 
= 2 e^{\vartheta} \frac{\Lambda^{\frac{1}{4}}}{\Lambda_{\text{u.v.}}^{\frac{5}{4}}} {\tilde{\ell}^{-\frac{1}{2}}}  { e^{-2\tilde{\ell}}}~.
%= -e^{\vartheta} \frac{\Lambda^{\frac{1}{4}}}{\left(\Lambda_{\text{u.v.}}\right)^{\frac{5}{4}}}  \frac{4}{\sqrt{\pi}}K_{\frac{1}{2}}(2\tilde{\ell})~. %\frac{Q}{2b}
\end{equation}
The above are monotonically decreasing functions in $\ell$, and we will identify the latter in terms of the Wheeler-DeWitt equation in section \ref{Wavefunctional perspective}. 
\newline\newline
The partition function $\mathcal{Z}_{D^2}[\tilde{\ell}]$ does not resemble a trace over a discretely spaced set of states in any obvious way. One modern interpretation of this, following ideas in \cite{Saad:2019lba,Anninos:2013nra} that take a Wigner-type perspective \cite{Wigner:1967qdh}, is to view  it as an ensemble average that washes out the underlying discrete nature of the Hilbert space. Indeed, from the point of view of the matrix model, one relates the disk partition function to the leading large-$N$ expectation value, or more precisely the genus-zero part of the double scaling limit (d.s.l.)  for the macroscopic loop operator
\begin{equation}
\mathcal{Z}_{D^2}[\ell] = \lim_{\text{d.s.l.}} \frac{1}{\mathcal{Z}_0} \int \dd M e^{-V_\alpha(M)} \,\text{Tr} \, e^{-M \ell}~.
\end{equation}
The role of $\mathcal{Z}_0$ is to normalise the probability distribution over the random Hamiltonians. We might then view the integral over the $N\times N$ Hermitean matrix $M$ as an specific averaging procedure over Hamiltonians $V_\alpha(M)$ acting on a $N$-dimensional Hilbert space. The trace $\text{Tr}$ is then viewed as a thermal trace over an $N$-dimensional Hilbert space. Again, the output is not quite an ordinary quantum mechanical quantity, but an averaged quantity. Notice that from this perspective the dimension of the Hilbert space tends to infinity at large $N$.
\newline\newline
{\textbf{Two boundaries.}} One can also consider the gravitational path integral on the annulus, $\mathcal{A}_2$, with two disconnected $S^1$ at the boundary. In this case, the output is a function of two boundary lengths, $\ell_1$ and $\ell_2$. We can view the expression as coding properties of the Lorentzian theory quantised on a spatial interval. In the simplest case where $\ell_1=\ell_2=\ell$, the result reads \cite{Ambjorn:1990ji,Moore:1991ag,martinec2003annular}
\begin{equation}
\mathcal{Z}_{\mathcal{A}_2}[\tilde{\ell}] = e^{-4\tilde{\ell}}~.
\end{equation}
We might view the above as a single thermally populated state at temperature $\ell$. However, more general expressions at $\ell_1 \neq \ell_2$ reveal otherwise. In pure gravity, a matrix model computation readily reveals  the following expression \cite{Ginsparg:1993is}
\begin{equation}\label{l1l2}
\mathcal{Z}_{\mathcal{A}_2}[\tilde{\ell}_1,\tilde{\ell}_2]  =   2 \times \frac{e^{-2\tilde{\ell}_1} \times e^{-2\tilde{\ell}_2}}{\sqrt{\frac{\tilde{\ell}_1}{\tilde{\ell}_2}}+\sqrt{\frac{\tilde{\ell}_2}{\tilde{\ell}_1}}}~,
\end{equation}
which does note take the structure of a trace on the spatial interval Hilbert space. This is to be contrasted with the annulus partition function of an ordinary conformal field theory on the annulus which is indeed a trace over the interval Hilbert space. The difference is that upon coupling to gravity we must also integrate over the size of the spatial integral in a diffeomorphism invariant way. This was computed explicitly in eqn. (27) of \cite{martinec2003annular}. 
\newline\newline
Rather than an explicit Hilbert space expression, a more modern perspective might again interpret (\ref{l1l2}) as the expectation value for two thermal partition functions of a quantum mechanical theory on Euclidean thermal cycles of size $\ell_1$ and $\ell_2$, whose Hamiltonian is drawn from the same ensemble. This is essentially a direct reading of the matrix model calculation
\begin{equation}
\mathcal{Z}_{\mathcal{A}_2}[\ell_1,\ell_2] = \lim_{\text{d.s.l.}} \frac{1}{\mathcal{Z}_0} \int [\dd M] e^{-V_\alpha(M)} \,\text{Tr} \, e^{-M \ell_1} \,\text{Tr} \, e^{-M \ell_2}~.
\end{equation}
The two quantum mechanical theories are coupled, from this perspective, because their Hamiltonian, $M$, is drawn from the same ensemble. Again, the price to pay is that we do not have a precise quantum mechanical description, even at the fundamental level.

\section{Wavefunctional perspective}\label{Wavefunctional perspective}

In this section, we explore the relation between single boundary gravitational path integrals and the large-$N$ limit of the matrix model. The approach we take follows that of Hartle and Hawking \cite{hartle1983wave} who argued that the output of these path integrals produces solutions to the Wheeler-DeWitt equation \cite{dewitt1967quantum}. Throughout this section, we will take the theory to reside on an $S^1$ spatial slice, such that the physical states are purely functionals of the proper length, $\ell$, of the spatial circle. 

\subsection{Wheeler-DeWitt equation}

The Wheeler-DeWitt equation is an alternative way to describe the canonical Hilbert space of the gravitational theory. One describes quantum states in terms of wavefunctionals of the data along the spatial slice. As noted in the previous section, upon going to the conformal gauge we must ensure that our physical states, $|\Psi\rangle$, are $PSL(2,\mathbb{C})$ invariant. In particular, physical states are annihilated by the Hamiltonian and momentum generators
\begin{equation}\label{constraints in terms of L_0 Lian-Zuckerman}
\left(L_0^{\text{tot}}+\tilde{L}_0^{\text{tot}}\right)\ket{\Psi}=0 \hspace{2.5mm}  \text{and} \hspace{2.5mm} \left(L_0^{\text{tot}}-\tilde{L}_0^{\text{tot}}\right)\ket{\Psi}=0~,
\end{equation}
which are the two-dimensional counterpart of the Wheeler-DeWitt equations. The $L_0^{\text{tot}}$ and $\tilde{L}_0^{\text{tot}}$ are the zero-modes of the Virasoro generators of the combined matter, Liouville, and ghost sectors.
\newline\newline
Our task is to express the Lian-Zuckerman states in terms of functionals $\Psi_\alpha(\ell)$ of the proper length 
\begin{equation}\label{physical boundary length}
\ell =\int_{S^1}d\xi    \sqrt{\tilde{h}} e^{b\varphi}~,
\end{equation}
of the spatial $S^1$ described by the coordinate $\xi\sim\xi+2\pi$. The label $\alpha$ further specifies which Lian-Zuckerman state $\Psi_\alpha$ is encoding --- the one associated to a Liouville vertex operator $\mathcal{O}_\alpha$. To further simplify matters, it is convenient to use the spatial diffeomorphisms along $S^1$ to go to a gauge where the spatial profile of the Liouville field is independent of the spatial coordinates. This choice further simplifies the Wheeler-DeWitt Hamiltonian constraint to an ordinary differential equation taking the specific form\footnote{A crucial difference between the Hamiltonian constraint (\ref{minisuperspace WdW Lian-Zuckerman}), and the one appearing for ordinary Einstein gravity is the sign of the gravitational kinetic term. In (\ref{minisuperspace WdW Lian-Zuckerman}) one has an ordinary sign for the kinetic term, whilst in the case of Einstein gravity it is the opposite Klein-Gordon like `wrong' sign \cite{dewitt1967quantum}. A more realistic Liouville theory that has the same sign as Einstein gravity is known as the timelike Liouville theory, as studied in \cite{Anninos:2024iwf}.}
\begin{equation}\label{minisuperspace WdW Lian-Zuckerman}
\left(-\left(\ell {\partial_\ell} \right)^2+4\Lambda \ell^2+\nu_{\alpha_t}^2\right)\Psi_{\alpha_t}(\ell)=0~, \hspace{5mm} \nu_{\alpha_t} \equiv \pm\frac{2}{b}\left(\alpha_t-\frac{Q}{2} \right)~,
\end{equation}
where we have rescaled $\Lambda\rightarrow \tfrac{\Lambda}{4 \pi b^2}$. We can readily solve (\ref{minisuperspace WdW Lian-Zuckerman}) in terms of a Bessel function. The solution that decays at large $\ell$ is given by
\begin{equation}\label{BesselK WF}
\Psi_{\alpha_t}(\ell) = \mathcal{N}_{\alpha_t} K_{\nu_{\alpha_t}}(2\sqrt{\Lambda}{\ell})~,
\end{equation}
where $\mathcal{N}_{\alpha_t}$ is a normalisation factor.
\newline\newline
For two-dimensional pure gravity, the allowed $\nu_{\alpha_t}$ follow directly from the BRST cohomology of Lian and Zuckerman, and read 
\begin{equation}\label{nu values in WdW}
\nu^+_{\alpha_t}= \frac{1}{2}|12t-1| \hspace{2.5mm} \text{and} \hspace{2.5mm} \nu^-_{\alpha_t} = \frac{1}{2}|12t+5|~,
\end{equation}
where $t\in \mathbb{Z}$. More succinctly, the allowed $\nu_{\alpha_t}$ run over all half-integers $n+\tfrac{1}{2}$ with $n\in\mathbb{N}$ and $n \neq 1 \,\text{mod}\,3$. The lowest lying value of the index $\nu_{\alpha_t}$, namely $\nu^+_{\alpha_{t=0}} = \frac{1}{2}$, corresponds to the state associated with an insertion of the Liouville vertex operator $\mathcal{O}_b$. 
From (\ref{Lian-Zuckerman grav dressing condition}) we find the Lian-Zuckerman weight $\Delta_{-}^{\text{LZ}}(0)=-1$. The index $\nu^-_{\alpha_{t=0}} = \frac{5}{2}$ corresponds to a vertex operator with $\alpha=0$, and yields a result different from that of the undecorated gravitational disk path integral (\ref{ZdiskU}) \cite{Moore:1991ir}. We expect this difference is due to a different treatment of residual gauge redundancies. The $\alpha=0$ operator is associated to the $t_2$ coupling of the string equation in (\ref{seqg}). As such, the closest object (\ref{ZdiskU}) to the original Hartle-Hawking wavefunction \cite{hartle1983wave} is not part of the Lian-Zuckerman cohomology. Nonetheless, we will relate the $\ell$-dependence of (\ref{ZdiskU}) to a linear combination of two Lian-Zuckerman states at a later stage. The remaining collection of Lian-Zuckerman operator insertions enter in a similar fashion. 
\newline\newline
We note that the solution (\ref{BesselK WF}) diverges as $\ell^{-|\nu_{\alpha_t}|}$ for small $\ell$. Such a small-size divergence is a characteristic feature of two-dimensional quantum gravity coupled to $c<1$ conformal matter at low topology. It contrasts the smooth behaviour at small spatial size of the Hartle-Hawking wavefunction \cite{hartle1983wave}. Relatedly, the  gravitational sphere path integral exhibits divergences for spheres of small physical area \cite{Zamolodchikov:1982vx,Muhlmann:2021clm}. The wavefunctions (\ref{BesselK WF})  are reproduced by the matrix model analysis in \cite{Moore:1991ir}.  One also has solutions to (\ref{BesselK WF}) that diverge exponentially at large $\ell$, but these are less physically meaningful given the infrared nature of their divergence.

\subsection{Disk partition function}

An alternative way to compute solutions to the Wheeler-DeWitt equation follows the path integral considerations of Hartle and Hawking \cite{hartle1983wave}. The basic reasoning mirrors the path integral construction of solutions to the Schr\"odinger equation in quantum mechanics and quantum field theory. 
\newline\newline
Constructing quantum states from a gravitational path integral is, generally, more subtle due to the conformal mode problem that renders the Euclidean path integral challenging to define \cite{gibbons1978path,polchinski1989phase}, as well as the absence of a clear resolution to the identity. The latter point stems from the fact that when we integrate over all metrics, we are in effect also integrating over all times, so it is less clear what one means by inserting a complete set of states at some intermediate time between an initial and final state. Further to this, the Hamiltonian constraint is not implemented in a sharp way from the path integral perspective. This is sensitive to the choice of contour for the lapse function, $N$, which a priori takes non-negative values --- it does not act as an ordinary Lagrange multiplier in the path integral  \cite{Teitelboim:1981ua,Teitelboim:1983fh}. One could suggest an alternative contour for $N$ to amend this, but it appears to come at the price of the path integral and matrix model pictures disagreeing.
\newline\newline
In any case, we will explore now the object most closely mimicking the quantum mechanical construction, which is a disk path integral with a gravitationally dressed operator insertion in its interior. Throughout this section, we will only consider the disk topology with an $S^1$ boundary of proper size $\ell$. Prior to any gauge fixing, the path integral of interest is
\begin{equation}\label{Zdisk}
\mathcal{Z}_{D^2}[\ell] = e^{\mathcal{\vartheta}} \int \frac{\mathcal{D}g_{ij}}{{\text{vol}} \,\mathcal{G}_{\text{diff}}} e^{-\Lambda \int d^2x\sqrt{g} -S_{\text{bdy}}} \, \mathcal{O}_{\text{grav}}(g_{ij})~.
\end{equation}
In order for the insertion $\mathcal{O}_{\text{grav}}$ to be gauge invariant, we must integrate its location over all points on the disk. A simple example of such an operator is the area operator $\mathcal{O}_{\text{grav}} = \int_{D^2} \dd^2x \sqrt{g}$, with suitable boundary conditions at the $S^1$ boundary. To render the path integral (\ref{Zdisk}) operationally meaningful, we again resort to the conformal gauge. There, it becomes a Liouville disk path integral \cite{fateev2000boundary, teschner2001liouville}, accompanied by the $\mathfrak{bc}$-ghost disk path integral. The gauge invariant operator $\mathcal{O}_{\text{grav}}$ becomes one of the Lian-Zuckerman operators analysed in the previous section. The Liouville disk path integral, $\mathcal{Z}^{(\alpha)}_{\text{L}}[\ell]$,  has been solved via bootstrap methods in \cite{fateev2000boundary,teschner2001liouville}. Upon inserting a vertex operator $\mathcal{O}_\alpha$, it yields the following expression
\begin{equation}
\mathcal{Z}^{(\alpha)}_{\text{L}}[\ell] \propto K_{\nu_{\alpha}}(2\sqrt{\Lambda}\ell)~. % \Lambda^{\nu_{\alpha}}
\end{equation}
The above is responsible for the entire $\ell$ dependence in $\mathcal{Z}_{D^2}[\ell]$, and agrees precisely with the solutions, $\Psi_\alpha(\ell)$ in (\ref{BesselK WF}), of the Wheeler-DeWitt equation once we select the appropriate values of $\alpha_t$.
\newline\newline
We note, again, that at face value we have found an infinite number of solutions to the Wheeler-DeWitt equation, one for each $\alpha_t$ appearing in the list of Lian-Zuckerman operators (\ref{Lian-Zuckerman operators}), paralleling the discussion in section \ref{infcouplings}. As before, to asses their physicality we need to resort to an inner product, now expressed in the language of Wheeler-DeWitt.

\subsection{Gravitational inner product} \label{section inner prdocut genus zero}

From the perspective of the Wheeler-DeWitt equation, what we are after is a bilinear pairing $(\Psi_{\alpha},\Psi_{\beta})$ that takes two wavefunctionals as  input. The inner product on the $\Psi_{\alpha}$ is always accompanied by an inner product on the $\mathfrak{bc}$-ghost Hilbert space also. From the perspective of Liouville theory as a quantum field theory, it is natural to consider the ordinary field theoretic pairing
\begin{equation}
(\Psi_{\alpha},\Psi_{\beta})_{\text{QFT}} \equiv \int \mathcal{D}\varphi_b \,\Psi_{\alpha}^*(\varphi_b) \Psi_{\beta}(\varphi_b)~,
\end{equation}
where $\varphi_b$ represents an arbitrary profile of the Liouville field along the spatial $S^1$. One issue with the above inner product is that yields at best $\delta$-function normaliseable eigenstates of the Hamiltonian, very much for the same reason that the Liouville two-point function  on $S^2$, (\ref{L2pt}), diverges. We previously identified such a divergence in terms of the volume of the dilatation subgroup of $PSL(2,\mathbb{C})$. Another issue, is that it most naturally applies to the physical spectrum of the quantum field theory that is built from vertex operators with $\alpha=\tfrac{Q}{2}+i P$ with $P\in\mathbb{R}$. But we have seen that the gravitational Hilbert space requires imaginary $P$.
\newline\newline
An inner product that might be a little closer to the underlying gravitational picture, that moreover renders the Wheeler-DeWitt Hamiltonian (\ref{minisuperspace WdW Lian-Zuckerman}) Hermitian, was suggested in \cite{seiberg1990notes, moore1991loops}, and reads
\begin{equation}\label{genus-zero inner product}
(\Psi_{\alpha},\Psi_{\beta})_{\text{grav}}  \overset{?}{\equiv} \frac{1}{\text{vol} \, \mathcal{G}_d}\int_{\mathbb{R}^+} \frac{d\ell}{\ell} \Psi^*_\alpha(\ell)\Psi_\beta(\ell)~.
\end{equation}
We note that the above inner product is expressed in a form that has fixed the spatial diffeomorphisms on $S^1$ to set $\varphi$ to a constant. As such we integrate only over the proper length $\ell$ of the $S^1$. Moreover, we must divide by the volume, $\text{vol} \, \mathcal{G}_d$, of the dilatation group to remove unfixed gauge redundancies. The Haar measure for the dilation group diverges logarithmically as
$\text{vol} \, \mathcal{G}_d=\int_{\mathbb{R}^+} \frac{d\gamma}{\gamma}$ which is logarithmically divergent due to the non-compact nature of the group. 
\newline\newline
Formally speaking, one might understand (\ref{genus-zero inner product}) as a group averaging projection \cite{Higuchi:1991tm} onto the dilation invariant part of the inner product. Indeed, the expression is formally invariant under rescalings $\ell \to \lambda \ell$, with $\lambda\in\mathbb{R}^+$. However, both the numerator and denominator generally diverge so we must still make sense of the expression. In order to do so, we can implement a gauge fixing procedure that further fixes the dilatation gauge symmetry. In effect, this would select a particular value of $\ell = \ell_0$ and render the inner product (\ref{genus-zero inner product}) finite.\footnote{The necessity of fixing the dilatation subgroup is somewhat reminiscent of the need to fix a cosmological time variable in order to compute probabilities in the Wheeler-DeWitt picture. A procedure that yields an inner-product independent of the gauge-fixing choice for $\ell$ would indicate the inner product is independent of the choice of time slice. } From a more invariant perspective  \cite{moore1991loops}, the dilatation invariant part of the integrand in (\ref{genus-zero inner product}) is obtained by expanding in a small-$\ell$ power series and selecting the constant term. This leads to the definition 
\begin{equation}\label{def genus-zero inner product}
(\Psi_{\alpha},\Psi_{\beta})_{\text{grav}}  \overset{?}{\equiv}  \left(   \Psi^*_{\alpha}(\ell)\Psi_{\beta}(\ell) \right)|_{\ell=0} ~.
%(\Psi_{\alpha},\Psi_{\beta})_{\text{grav}}  \overset{?}{\equiv} \partial_\ell \left(   \Psi^*_{\alpha}(\ell)\Psi_{\beta}(\ell) \right)|_{\ell=0} ~.
\end{equation}
Unfortunately, the above expression is still formal. The
%derivative for
physical wavefunctions (\ref{BesselK WF}) diverge at $\ell = 0$, and must be regularized. Operationally, we might  take the definition (\ref{def genus-zero inner product}) to be that we extract the constant part of the small-$\ell$ expansion of the integrand of (\ref{genus-zero inner product}). Doing so will yield a finite quantity for (\ref{def genus-zero inner product}). Again, nothing guarantees this inner-product will be positive. For instance, $(\Psi_{\nu^+_{0}},\Psi_{\nu^+_{0}})_{\text{grav}} = -|\mathcal{N}_{\nu_0^+}|^2\pi$. This echoes our discussion near (\ref{2ptRR}). More generally for non-vanishing $t$, we find
\begin{equation}
    (\Psi_{\nu^\pm_{t}},\Psi_{\nu^\pm_{t}})_{\text{grav}} = \pm \text{sign}\,t|\mathcal{N}_{\nu_t^\pm}|^2 \frac{\pi}{2\nu_t^\pm}~.
    %\quad    (\Psi_{\nu^-_{t}},\Psi_{\nu^-_{t}})_{\text{grav}} \equiv -\text{sign}\,t|\mathcal{N}_{\nu_t^-}|^2 \frac{\pi}{2\nu_t^-}~.
\end{equation}  
Perhaps we should view this as a regularized piece of a divergently positive part of the norm. We also note that with respect to the inner product (\ref{genus-zero inner product}), understood in this regulated sense, the $\Psi_{\nu^\pm_{t}}(\ell)$ for different values of $|\nu_t|$ are orthogonal.

\section{Higher genus contributions}\label{hgwf}

In this section, we  compute higher genus corrections to the gravitational path integral on a manifold with a single boundary. From the perspective of our discussion, we have two reasons for doing so. Firstly, from the perspective of Hartle and Hawking \cite{hartle1983wave}, nothing impedes us from considering these as potential contributions to their gravitational wavefunction. Indeed,  they are compatible with the type of no-boundary path integral Hartle and Hawking proposed to compute. Secondly, from the matrix model perspective, these contributions are there by construction, necessitated to account for  subleading terms in the planar expansion. 
%What is not guaranteed, and in fact does not occur, is that these contributions will yield a function of $\ell$ that solves the Hamiltonian constraint of the Wheeler-DeWitt equation (\ref{minisuperspace WdW Lian-Zuckerman}). 

\subsection{General procedure}

From a gravitational path integral perspective, we would have to compute the Liouville and $\mathfrak{bc}$-ghost path integral on a disk with handles. This is a somewhat daunting task we will not pursue here. Rather, we will proceed by computing this quantity from the matrix model side. We analyze the even simpler case of topological gravity \cite{witten1993two} in appendix \ref{topgrav}, which yields qualitatively similar but more explicit results. 
\newline\newline
The general procedure that we will follow is detailed in \cite{Banks:1989df}. (An alternative method implements the ideas of topological recursion, as studied in \cite{mertens2021liouville,Johnson:2026twg,johnson2026non}.) In the double scaling limit, one lands on an effective quantum mechanical problem governed by a single-particle Hamiltonian 
\begin{equation}
H_\kappa = -\frac{\kappa^2}{2} \partial^2_z + V_{\kappa}(z)~,
\end{equation}
with $z\in\mathbb{R}$ representing the position of the particle. We will further set $\hbar=1$. From this perspective, we see the emergence of  yet another Hilbert space --- that of a quantum mechanical theory whose Hamiltonian has a continuous spectrum. The potential $V_{\kappa}(z) = - \frac{1}{2} \mathcal{P}_{\kappa}(z)$ is obtained by solving the string equation  \cite{brezin1990exactly,gross1990nonperturbative,douglas1990strings}, which for pure two-dimensional quantum gravity is a Painlev\'e I equation of the form
\begin{equation}\label{Painlevé eq}
\mathcal{P}_\kappa(z)^2+\frac{\kappa^2}{3} \partial_z^2 \mathcal{P}_\kappa(z) - z = 0~.
\end{equation}
The parameter $\kappa$ can be scaled out, but we keep it as a bookkeeping device for the genus expansion. For instance, the disk topology goes as $\kappa^{-1}$. We are interested in the small-$\kappa$ expansion (or equivalently the large-$z$ expansion) which can be viewed as a WKB type limit from the quantum mechanical perspective. Upon expanding $\mathcal{P}_\kappa(z)$ in a $\kappa$-expansion
\begin{equation}\label{series solution Painlevé I}
\mathcal{P}_\kappa(z) = \sum_{n=0}^{\infty} \,  {a_n }{z^{\frac{1-5n}{2}}}\kappa^{2n}, % \hspace{5mm}
\end{equation}
we select the $a_0=-1$ branch to make contact with the matrix model. The coefficients $a_n$ satisfy the recursion relation (see \cite{ginsparg1991large,anninos2020notes} for reviews)
\begin{equation}\label{recusrion realtion Painlevé I}
a_{n+1} = \frac{25 n^2-1}{24}a_n+\frac{1}{2}\sum_{m=1}^n a_m a_{n+1-m}~, \hspace{5mm} a_0=-1~, \hspace{5mm} n\geq0~.
\end{equation}
To the first few orders in the small $\kappa$ expansion, we have
\begin{equation}\label{solution Painlevé}
\mathcal{P}_\kappa(z) = -{z}^{\frac{1}{2}}+\frac{1}{24}z^{-2}\kappa^2+\frac{49}{1152}z^{-\frac{9}{2}}\kappa^4+\frac{1225}{6912} z^{-7}\kappa^6 + \ldots
%O\left(\kappa ^8\right)
%+\frac{4412401}{2654208 z^{\frac{19}{2}}}\kappa^8+\sum_{n=5}^{\infty}a_n\, \frac{\kappa^{2n}}{z^{\frac{5n-1}{2}}}.
\end{equation}
The solution for $\mathcal{P}_\kappa(z)$  yields a growing potential, $V_\kappa(z)$, in addition to being the choice that matches the matrix integral (for a review, see \cite{anninos2020notes}). The quantum mechanical quantity we are interested in is the partial trace
\begin{equation}\label{Wds}
\mathcal{W}_\kappa(\ell) = \int_{\Lambda}^\infty \text{d}z_0  \braket{z_0|e^{-\ell  H_\kappa}|z_0}~.
\end{equation}
This is the expectation value of the macroscopic loop operator of the matrix model, as computed in the double scaling limit. It is often convenient to scale out the $\Lambda$ dependence by taking $z= \Lambda \tilde{z}$, $\ell =  \Lambda^{-\frac{1}{2}}\tilde{\ell}$, and $\kappa=\Lambda^{\frac{5}{4}}\tilde{\kappa}$. We will often present results in the $\Lambda=1$ frame, and we use the tilded variables for this. A simple rescaling reinstates the dependence on $\Lambda$. From the gravitational path integral perspective, we are placing the theory on a manifold with a boundary of proper length $\ell$. The terms in the small-$\kappa$ expansion compute contributions of increasing genus, with $\kappa^{-1}$ capturing the genus-zero case, and so on. 

\subsection{Methodology}\label{Small ell behavior}

One method to compute $\mathcal{W}_\kappa(\ell)$, is via a quantum mechanical Euclidean path integral. The Euclidean action is
\begin{equation}\label{Euclidean action}
{S}_{{E}}=\int_0^{\ell}\,\text{d}\tau \left( \frac{1}{2 \kappa^2}(\partial_\tau{z})^2+V_\kappa(z) \right)~,
\end{equation}
and we consider paths that begin and end at $z_0$. When working in the tilded variables, we must also take $\tau = \Lambda^{-\frac{1}{2}} \tilde{\tau}$. From the path integral perspective, we have
\begin{equation}\label{path integral heat kernel}
\bra{z_0}e^{-\ell  H}\ket{z_0} = \int\mathcal{D}z\, e^{-{S}_{{E}}[z(\tau)]}~.
\end{equation}
We can treat the problem semiclassically, in a small-$\kappa$ expansion. Details can be found in appendix \ref{Heat kernel calc}. To leading order, we have a classical path $z_{\text{cl}}(\tau)$, itself admitting an expansion in small $\kappa$, governed by the potential $V_\kappa(z)$. On top of this are small quantum mechanical fluctuations $\delta z(\tau) \equiv \kappa\,\zeta(\tau)$ where $\zeta(0)=\zeta(\ell)=0$.
%WKB
\newline\newline
A complementary method is to systematically analyse the small-$\ell$ behaviour of our path integral (\ref{path integral heat kernel}) employing a Schwinger-DeWitt type expansion. We will compute the propagator, $K(z_1,z_2;\ell)$, between two points. To leading order we have the propagator of a free particle. For our problem, this is given by
\begin{equation}
K^{(0)}(z_1,z_2;\ell) \equiv \frac{e^{ -\frac{(z_1-z_2)^2}{2 \kappa ^2 \ell}}}{\sqrt{2 \pi \kappa^2  \ell}}~.  %K(z_1,z_2;\ell) \approx 
\end{equation}
The small-$\ell$ expansion then takes the general form
\begin{equation}
K(z_1,z_2;\ell) =  K^{(0)}(z_1,z_2;\ell)\left(1 + a_1(z_1,z_2) \ell  + a_2(z_1,z_2) \ell^2 + \ldots \right)~,
\end{equation}
where the coefficients $a_n(z_1,z_2)$ are systematically calculable for an arbitrary potential $V(z)$. Indeed, from the heat equation $\partial_\ell K(z_1,z_2;\ell) = -H_\kappa K(z_1,z_2;\ell)$ we arrive at a recursive definition for the coefficients $a_n(z_1,z_2)$
\begin{equation}
    \left(n+(z_1-z_2)\partial_{z_1}\right)a_n(z_1,z_2) = \frac{\kappa^2}{2}\partial_{z_1}^2a_{n-1}(z_1,z_2)-V_\kappa(z_1)a_{n-1}(z_1,z_2)~.
\end{equation}
In particular, 
\begin{equation}
    a_n(z_1,z_2) = \int_0^1 ds s^{n-1}\left(\frac{\kappa^2}{2}\frac{\partial^2}{\partial x(s)^2}-V_{\kappa}\left(x(s)\right)\right)a_{n-1}(x(s),z_2)~,
\end{equation}
where $x(s)=z_2+s(z_1-z_2)$ and $a_0(z_1,z_2)=1$. For a general potential, the first two read
\begin{eqnarray}
a_1(z_1,z_2) &=& - \int_0^1ds_1 V_\kappa\left(z(s_1)\right)~, \\
a_2(z_1,z_2) &=& -\int_0^1ds_2s_2\int_0^1 ds_1\left(\frac{\kappa^2}{2}\frac{\partial^2}{\partial x_2^2}-V_\kappa(x_2)\right)V_\kappa(x_1)~,
\end{eqnarray}
where $x_i(s_i) =z_2+(x_{i+1}-z_2)s_i$ and $x_{3}=z_1$. Derivatives with respect to different $x_i$ are related via
\begin{equation}
    \frac{\partial}{\partial x_i}= s_{i-1}\frac{\partial}{\partial x_{i-1}}~.
\end{equation}
For the potential given by (\ref{Painlevé eq}) we have % ({\color{blue} to be corrected})
\begin{eqnarray}
a_1(z_1,z_2) &=&  -\frac{z_1+\sqrt{z_1z_2}+z_2}{3(\sqrt{z_1}+\sqrt{z_2})}+\frac{\kappa^2}{48 z_1z_2}+\mathcal{O}(\kappa^4) ~, \\ %- \int_0^1 \text{d}s \,V(z_1+s(z_2-z_1))~,  \\
a_2(z_1,z_2) &=& \frac{(z_1+\sqrt{z_1z_2}+z_2)^2}{18(\sqrt{z_1}+\sqrt{z_2})^2}-\frac{\kappa^2}{144}\frac{(\sqrt{z_1}-\sqrt{z_2})^2(z_1+5\sqrt{z_1z_2}+z_2)}{z_1(\sqrt{z_1}+\sqrt{z_2})^3z_2}+\mathcal{O}(\kappa^4)~. %\frac{1}{2} a_1(z_1,z_2)^2 - \frac{\kappa^2}{2} \int_{0}^{1}  \text{d}s\, s(1-s)V''(z_1+s(z_2-z_1))~.
\end{eqnarray}
We further note that the coincident point $z_1=z_2$ limit of these is given by
\begin{eqnarray}
a_1(z,z) &=& -V_\kappa(z)~, \\
\quad a_2(z,z) &=& -\frac{\kappa^2}{12}V_\kappa''(z)+\frac{1}{2}V_\kappa^2(z)~, \\ 
\quad a_3(z,z) &=& -\frac{\kappa^4}{240}V_\kappa^{(4)}(z)+\frac{\kappa^2}{24}V_\kappa'(z)^2+\frac{\kappa^2}{12}V_\kappa(z)V_\kappa''(z)-\frac{1}{6}V_\kappa^3(z)~.
\end{eqnarray}
The above agree with the appropriately rescaled diagonal heat kernel coefficients reported in the mathematics literature \cite{hitrik2003regularized,pushnitski2006high}, and one can proceed systematically to higher order. A resummed version, that keeps all the contributions from $a_1(z_1,z_2)$ at $\kappa=0$ in the exponent, is given by %(\textcolor{blue}{to be corrected})
% \begin{equation}
% K(z_1,z_2;\ell) =  K^{(0)}(z_1,z_2;\ell)e^{a_1(z_1,z_2)\ell}\left(1  + \frac{\kappa^2\ell^2}{12}\frac{1}{(\sqrt{z_1}+\sqrt{z_2})^3} -\frac{\kappa^4\ell^2}{4608z_1^2z_2^2} \ell^2 + \ldots \right)~.
% %K(z_1,z_2;\ell) =  K^{(0)}(z_1,z_2;\ell)e^{-a_1(z_1,z_2)\ell}\left(1  - \frac{\kappa^2\ell^2}{2} \int_0^1 \,s(1-s) \text{d}s V''(z_1+s(z_2-z_1))  + \ldots \right)~.
% \end{equation}
% At coincidence $z_1=z_2=z_0$, this reads
\begin{equation}
K(z,z;\ell) = \frac{1}{\sqrt{2\pi\kappa^2\ell}}e^{\left(-\frac{\sqrt{z}}{2}+\frac{\kappa^2}{48z^2}+\ldots\right)\ell}\left(1  + \frac{\kappa^2\ell^2}{96}{z^{-\frac{3}{2}}}  +\frac{\kappa ^2 \ell^3}{384} z^{-1} + \ldots \right)~.
\end{equation}
%\frac{49 \kappa^4}{2304 z_0^{9/2}}, + \frac{\kappa^4 \ell^2}{96} z_0^{-4}
Expanding the above expression at small-$\kappa$, and integrating against $z$ is an alternative method to obtain the genus contributions. What is more, given the general heat kernel coefficients $a_n(z,z)$ reported in \cite{hitrik2003regularized,pushnitski2006high}, along with the recursive solution (\ref{recusrion realtion Painlevé I}), we can hope to have a systematic and efficient construction of the genus corrections to arbitrary order. This is left for future work.

\subsection{Low genus expressions}

We can now implement the procedure to systematically compute the first few genus corrections. In the case of the semiclassical approach, we evaluate the on-shell action for the classical path governed by $V_\kappa(z)$, and the corresponding one-loop determinant of the corrected differential operator $\mathcal{O}_\kappa$.  The latter is computed using the method of Gelfand and Yaglom \cite{gel1960integration}. Higher-loop diagrammatic contributions are only necessary at genus two and higher. Although we spare the reader most details, we emphasise that the on-shell actions we have computed indeed satisfy the Hamilton-Jacobi equation, and we will test our expressions against small-time heat kernel type expansions in what follows. When using the heat kernel method, we check that the off-diagonal heat kernel elements obey the heat kernel equation, and cross check the two methods when possible.
\newline\newline
{\textbf{Genus-zero.}} To leading order, corresponding to genus zero, we have a completely undisturbed classical path $z_{\text{cl}}(\tau) = z_0$, yielding the leading unit norm approximation %({\color{blue}fix to our conventions})
\begin{equation}\label{genus zero path integral heat kernel}
\mathcal{W}^{(0)}(\tilde{\ell}) =\int^\infty_{1}  \text{d}\tilde{z}_0 \,\bra{\tilde{z}_0}e^{-\tilde{\ell}  \tilde{H}_\kappa}\ket{\tilde{z}_0} \approx \frac{1}{\tilde{\kappa}}\left(2\tilde{\ell}^{-\frac{5}{2}}+\tilde{\ell}^{-\frac{3}{2}}\right)e^{-\frac{\tilde{\ell}}{2}}~,
\end{equation}
up to an overall normalisation which we have chosen for future convenience. Once this normalisation is fixed, we cannot choose it independently for the higher genus expressions. We note that (\ref{genus zero path integral heat kernel}) is equivalent to (\ref{ZdiskU}).  To calculate the above, we must also take into account the one-loop determinant of the kinetic operator $\mathcal{O}_\kappa = -\partial_{\tilde{\tau}}^2$. The $\ell$-dependence in (\ref{genus zero path integral heat kernel}) agrees with the linear combination of two wavefunctions (\ref{BesselK WF})  with $\nu^+_{t=0} = \frac{1}{2}$ and $\nu^-_{t=0}=\frac{5}{2}$ once we rescale $\ell\to4\ell$, which is simply a field redefinition.  Expressed in terms of Bessel functions, the genus zero contribution reads
\begin{equation}\label{W0}
   \mathcal{W}^{(0)}(\tilde{\ell}) = \frac{1}{6\sqrt{\pi} \tilde{\kappa}}\left(K_{\frac{5}{2}}\left( \tfrac{\tilde{\ell}}{2}\right)-K_{\frac{1}{2}}\left(\tfrac{\tilde{\ell}}{2}\right)\right)~.
   %\frac{\sqrt{2}}{3\pi \tilde{\kappa}}\left(K_{{5}/{2}}( \tilde{\ell}/2)-K_{{1}/{2}}(\tilde{\ell}/2)\right)~.
\end{equation}
We will also be able to express the higher genus results in terms of such Bessel functions. The indices, $\tfrac{1}{2}$ and $\tfrac{5}{2}$ are compatible with the the two Lian-Zuckerman states (\ref{BesselK WF}) at $t=0$. In other words, we might cautiously interpret (\ref{W0}) as a solution to the Wheeler-DeWitt equation given by a particular linear combination thereof. More general genus-zero wavefunctions can be obtained by turning on additional couplings in the string equation (\ref{seqg}) \cite{Moore:1991ir,Belavin:2010ba}. For instance the $t_2$ coupling yields for genus-zero wavefunctions with a non-analytic structure compatible with a $\nu=\tfrac{5}{2}$ Bessel $K$ function.
\newline\newline
{\textbf{Genus-one.}} Things become a little more interesting at genus one. Here, the classical path becomes slightly perturbed and the functional determinant has an $\mathcal{O}(\kappa^2)$ correction that must be taken into account, as detailed in appendix \ref{appPI}. Putting it all together, and going to the $\Lambda=1$ frame, we have the genus-one contribution\footnote{We remark that inserting three area operators, (\ref{Area operator}), into the genus-zero contribution, (\ref{genus zero path integral heat kernel}), produce the genus-one contribution (\ref{The wavefunction genus one}) up to an $\ell$-independent overall pre-factor.}
\begin{equation}\label{The wavefunction genus one}
\mathcal{W}^{(1)}(\tilde{\ell}) = \frac{\tilde{\kappa}}{384}\left(2\tilde{\ell}^{\frac{1}{2}}+\tilde{\ell}^{\frac{3}{2}}\right)e^{-\frac{\tilde{\ell}}{2}}~.
%\frac{\tilde{\kappa}}{96\sqrt{2\pi}}\left(2\tilde{\ell}^{1/2}+\tilde{\ell}^{3/2}\right)e^{-\tilde{\ell}/2}~.
%=  \frac{\tilde{\kappa}}{24 \sqrt{2 \pi }} \left(\tilde{\ell}^{1/2}+2 \tilde{\ell}^{3/2}\right) e^{-2\tilde{\ell}}.
%  \frac{1}{\tilde{\kappa}}\left(\tilde{\ell}^{-5/2}+\tilde{\ell}^{-3/2}\right) e^{-\tilde{\ell}} +
%&+\frac{\Lambda^{\frac{1}{4}}\sqrt{l} e^{-2 l \sqrt{\Lambda }} \left(20 l^4 \Lambda ^2+58 l^3 \Lambda ^{3/2}+84 l^2 \Lambda +70 l \sqrt{\Lambda }+35\right)}{2880 \sqrt{2 \pi } \Lambda ^{15/4}}\kappa ^3 +\mathcal{O}(\kappa^5).
\end{equation}
Expressed in terms of Bessel functions, the genus one contribution reads
\begin{equation}%\label{The wavefunctions genus-zero, one...}
\mathcal{W}^{(1)}(\tilde{\ell}) = \frac{\tilde{\kappa}}{2304 \sqrt{\pi} }  \tilde{\ell}^3\left(K_{\frac{5}{2}}\left( \tfrac{\tilde{\ell}}{2}\right)-K_{\frac{1}{2}}\left(\tfrac{\tilde{\ell}}{2}\right)\right) ~.
% %\frac{\tilde{\kappa}}{576 \sqrt{2} \pi }  \tilde{\ell}^3\left(K_{{5}/{2}}( \tilde{\ell}/2)-K_{{1}/{2}}(\tilde{\ell}/2)\right) ~.
% \end{equation}
% \mathcal{W}(\tilde{\ell}) \approx  \frac{\sqrt{2}}{3\pi\tilde{\kappa}}\left( K_{{5}/{2}}(\tilde{\ell}/2)-  K_{{1}/{2}}(\tilde{\ell}/2)\right) +  + \ldots~.
% %\frac{\sqrt{2}}{3\pi\tilde{\kappa}}\left( K_{{5}/{2}}(2 \tilde{\ell})-  K_{{1}/{2}}(2 \tilde{\ell} )\right) + \frac{\tilde{\kappa}}{9 \sqrt{2} \pi }  \tilde{\ell}^3\left(K_{{5}/{2}}(2 \tilde{\ell})-K_{{1}/{2}}(2 \tilde{\ell})\right)  + \ldots~.
% % \\
% %          \frac{ l^5 }{90 \sqrt{2} \pi  \Lambda ^{5/4}}\left(K_{\frac{1}{2}}\left(2 \sqrt{\Lambda} l\right)-\frac{25}{12} K_{\frac{3}{2}}\left(2 \sqrt{\Lambda } l\right)+2 K_{\frac{5}{2}}\left(2 \sqrt{\Lambda} l\right)+\frac{1}{3} K_{\frac{9}{2}}\left(2 \sqrt{\Lambda} l\right)\right)\kappa ^3 +\mathcal{O}(\kappa^5)~.
\end{equation}
We can compare the above expression to an integral representation for the same quantity that follows from the topological recursion relation and specifically eqn. (7.47) of \cite{mertens2021liouville}, which we interpret as follows %{\color{blue} xxx fix}
\begin{equation}
\mathcal{W}_{\text{MT}}^{(1)}(\tilde{\ell}) = \tilde{\kappa} \int_{\mathbb{R}^+} d\lambda \, \lambda \, \text{tanh}(\pi\lambda)K_{i\lambda}(\tilde{\ell})\left(\frac{\pi^2}{12}+ \lambda^2 \frac{\pi^2}{27}\right)~. %3 p^2
\end{equation}
Numerically computing the above integral yields good agreement with our result (\ref{The wavefunction genus one}).  Moreover, we can compare the expression with the inverse Laplace transform of eqn. (B.1) in \cite{Johnson:2026twg} for $g=1$, and again find agreement once we appropriately rescale the parameters.
\newline\newline
{\textbf{Genus-two.}} At genus two, in addition to the corrected determinant and classical path, we need to take into account a diagrammatic correction. The details of the computation are summarised in appendix \ref{appPI}. The resulting expression reads
\begin{equation}\label{Genus two wavefunction}
\mathcal{W}^{(2)}(\tilde{\ell}) = \frac{\tilde{\kappa}^3}{1474560}\left(2240\tilde{\ell}^{\frac{1}{2}}+1120\tilde{\ell}^{\frac{3}{2}}+336\tilde{\ell}^{\frac{5}{2}}+58 \tilde{\ell}^{\frac{7}{2}}+5\tilde{\ell}^{\frac{9}{2}}\right) e^{-\frac{\tilde{\ell}}{2}}~.
%\tilde{\kappa}^{3}\left(\frac{7}{2304}\tilde{\ell}^{\frac{1}{2}}+ \frac{7}{1152}\tilde{\ell}^{\frac{3}{2}} + \frac{7}{960}\tilde{\ell}^{\frac{5}{2}} + \frac{29}{5760} \tilde{\ell}^{\frac{7}{2}} + \frac{1}{576} \tilde{\ell}^{\frac{9}{2}} \right)e^{-2\tilde{\ell}}~.
%\frac{\tilde{\kappa}^3}{1474560}\left(2240\tilde{\ell}^{\frac{1}{2}}+1120\tilde{\ell}^{\frac{3}{2}}+336\tilde{\ell}^{\frac{5}{2}}+58 \tilde{\ell}^{\frac{7}{2}}+5\tilde{\ell}^{\frac{9}{2}}\right) e^{-2{\tilde{\ell}}}~.
%\frac{\tilde{\kappa}^3}{2880\sqrt{2\pi}}\left(20\tilde{\ell}^{9/2}+58 \tilde{\ell}^{7/2}+84\tilde{\ell}^{5/2}+70\tilde{\ell}^{3/2}+35\tilde{\ell}^{1/2} \right) e^{-2\tilde{\ell}}~.
\end{equation}
We note again the exponential decay in $\tilde{\ell}$. Again, the result can be expressed in terms of Bessel-$K$ functions, now with an overall $\tilde{\ell}^5$ multiplicative factor. We can again compare our result (\ref{Genus two wavefunction}) with the inverse Laplace transform of eqn. (B.1) in \cite{Johnson:2026twg} for $g=2$ and find agreement once we appropriately rescale the parameters.
\newline\newline
{\textbf{Genus-three to six.}} At genus-three and above, we have found it more efficient to use the Euclidean heat kernel method. We have obtained explicit results up to genus-six, and report them in appendix \ref{Heat kernel calc 2}. We successfully compare the expressions for genus three with existing expressions in the literature which have been computed through different means \cite{Johnson:2026twg}. The expressions for genus-four and above, namely (\ref{Genus four wavefunction}), (\ref{Genus five wavefunction}), and (\ref{Genus six wavefunction}),  appear to be new.  
\newline\newline
{\textbf{Genus-$h$ structure.}} Although we stopped at genus-six, our heat kernel method is fairly efficient in proceeding to higher-genus. We anticipate that general structure for the genus-$h$ contribution, at $h\geq1$,  takes following form 
\begin{equation}\label{Wh}
\mathcal{W}^{(h)}(\tilde{\ell}) = \tilde{\kappa}^{2h-1} e^{-\tfrac{\tilde{\ell}}{2}}\sum_{n=0}^{3h-2} \zeta_{n}\,\tilde{\ell}^{n+\frac{1}{2}}~.
%\mathcal{W}^{(h)}(\tilde{\ell}) = p_h(\tilde{\ell}) e^{-\tilde{\ell}/2}~, %\quad\quad p_h(\tilde{\ell}) =  \left(a_h\tilde{\ell}^{-5/2} + b_h \tilde{\ell}^{-3/2} \right)~.
\end{equation}
%where $\pi_n$ are some constant rational-valued coefficients. 
Up to an overall factor, it follows from (\ref{recusrion realtion Painlevé I}) and the general structure of heat kernel contributions \cite{hitrik2003regularized,pushnitski2006high} that the coefficients appearing in $\zeta_n$ are rational. It would be interesting to obtain an exact expression, including all coefficients, in a more explicit form. Interestingly, for all $h\ge1$ the wavefunctions are well-behaved in the $\tilde{\ell}\to0^+$ limit, in line with the picture of Hartle-Hawking. This contrasts the genus-zero case.

\subsection{Large-$\ell$ behavior}\label{largelsec}

So far we have considered the problem order by order in a small-$\kappa$ expansion. It is also of interest to obtain results for the full expansion. In particular, we can anticipate the large-$\ell$ limit of $\mathcal{W}_\kappa(\ell)$ in (\ref{Wds}). In an ordinary quantum system, the large Euclidean time limit drives the system to its ground state. For a sufficiently smooth state, $|\psi\rangle$, one has $\langle\psi|e^{-\ell H}|\psi\rangle \approx e^{-\ell E^{(\kappa)}_0} |\langle E^{(\kappa)}_0 |\psi\rangle|^2$ with $|E^{(\kappa)}_0\rangle$ being the lowest energy eigenstate with ground state energy $E^{(\kappa)}_0$, itself a function of $\kappa$. The situation at hand is a little more complicated because we do not have complete knowledge of the energy eigenstates. Moreover, the potential $V_\kappa(z)$  associated to $\mathcal{P}_\kappa(z)$ in (\ref{Painlevé eq}) yields a continuous spectrum of scattering states. However, there is some numerical evidence \cite{Banks:1989df} that the more complete form of the potential $V_\kappa(z)$ has a minimum hence admitting a discrete set of energy eigenstates. 
\newline\newline
The exponential behaviour is in agreement with the perturbative results which all decay exponentially in $\ell$. What is not captured by the perturbative results is that the exponent is $\kappa$-dependent. It is anticipated, however, by the fact that the subleading pieces of higher genus contributions grow increasingly fast. For instance, taking the expression (\ref{Wds}}) up to genus one, we see that when $\tilde{\ell} \tilde{\kappa}^{\frac{2}{3}}\gg 1$, the perturbative approximation breaks down. Indeed, in this deep infrared regime, we must resum terms from all genus contributions such that they yield the corrected exponential decay.\footnote{We remark here that although the large ${\ell}$ behaviour, $e^{-\ell E^{(\kappa)}_0}$, seems fairly innocuous, it could have more significant implications in a slightly modified setting. For instance, in timelike Liouville theory  \cite{Anninos:2024iwf}, the wavefunctions at genus zero oscillate at large $\tilde{\ell}$. If, for whatever reason, higher genus corrections modify the exponent with a complex coefficient, oscillations can turn into exponentially growing/decaying effects.} Finally, granting the general form (\ref{Wh}), we note the limiting behaviour
\begin{equation}
\lim_{\tilde{\ell}\to\infty} \tilde{\kappa}^{-2}\tilde{\ell}^{-3} \frac{\mathcal{W}^{(h+1)}(\tilde{\ell})}{\mathcal{W}^{(h)}(\tilde{\ell})} =  \alpha_h~,
\end{equation}
with $\alpha_h$ some $h$-dependent coefficient. In our conventions, we can read off from our results in appendix \ref{Heat kernel calc 2} that $\alpha_0 = \tfrac{1}{384}$, $\alpha_1 = \tfrac{1}{768}$, $\alpha_2 = \tfrac{1}{1152}$, $\alpha_3 = \tfrac{1}{1536}$, $\alpha_4 = \tfrac{1}{1920}$, and $\alpha_5 = \tfrac{1}{2304}$. We are tempted to postulate $\alpha_h = \tfrac{1}{384(1+h)}$, but we reserve the question of $\alpha_h$ to an analysis of its own \cite{sam}. An alternative method employs the recursional methods of \cite{Ambjorn:1992gw,Eynard:2007kz,mirzakhani2007simple,Belavin:2010bs,Johnson:2026twg}. We further note that eqn. (20) of \cite{Giacchetto:2026gpq}, and the ensuing methods of that paper, are particularly relevant to this question.
%Indeed, using the inverse Laplace transform of (B.2), (B.3) and (B.4) in \cite{Johnson:2026twg}, we find agreement with the proposed $\alpha_h$.
\newline\newline
The relative $\tilde{\ell}^3$ as well as the $h$-dependence of $\alpha_h$ also appears in a simple example of pure topological gravity studied in appendix \ref{topgrav}. We find $\alpha_h = \tfrac{1}{96(1+h)}$ for the model in appendix \ref{topgrav}. This suggests there is a topological explanation the large-$\tilde{\ell}$ behaviour, likely related to the fact that at large $\tilde{\ell}$ we are probing the edge point of the eigenvalue distribution in the matrix model where the behavior is more universal. Indeed, a similar pattern appears in the Laplace transformed expressions $\mathcal{W}^{(h)}(\mu_B)$ shown, for instance, in appendix B of \cite{Johnson:2026twg} for which one anticipates the genus-$h$ leading non-analytic behaviour. Adjusting to the convention of this section, the leading non-analyticities reported in \cite{Johnson:2026twg} take the form $\mathcal{W}^{(h)}(\mu_B) \propto   \left( \sqrt{\Lambda}+2\mu_B \right)^{\frac{1}{2}-3h}$. Upon performing an inverse Laplace transform from $\mu_B$ to $\tilde{\ell}$, these indeed convert to terms that take the form $\mathcal{W}^{(h)}(\tilde{\ell}) \propto e^{-\frac{\tilde{\ell}}{2}} \,\tilde{\ell}^{3h-\frac{3}{2}} =  e^{-\frac{\tilde{\ell}}{2}} \,\tilde{\ell}^{-\frac{3}{2}\chi_{h,1}}$.

\subsection{Non-perturbative effects}

As a final remark, it is important to note that the perturbative solution (\ref{solution Painlevé}) of (\ref{Painlevé eq}) is not Borel summable \cite{gross1990nonperturbative}. In particular, the correct solution has additional non-perturbative contributions which at small $\kappa$ take the form \cite{shenker1991strength,ginsparg1991large}
\begin{equation}\label{nonpert}
\delta \mathcal{P}_\kappa(z) \approx \varepsilon_\kappa\left({z^{\frac{5}{4}}{\kappa^{-1}}}\right)^{-\frac{1}{10}} \exp \left(-\gamma  {z^{\frac{5}{4}}}{\kappa^{-1}} \right)~.
%\varepsilon_\kappa\left({\kappa^{\frac{4}{5}}}{z^{-1}}\right)^{\frac{1}{8}} \exp \left(-\gamma  {z^{\frac{4}{5}}}{\kappa^{-1}} \right)~.
\end{equation} 
For the case of pure two-dimensional gravity that we are interested in, $\gamma$ is real valued (see, for example, section 4 of \cite{anninos2020notes}), but this is not the case for more general theories of gravity coupled to minimal models where $\gamma$ can also be complex. Moreover, nothing precludes the coefficient, $\varepsilon_\kappa$, from being complex valued \cite{Hanada:2004im}. These terms will appear as small corrections to the potential $V_\kappa(z)$. For $\gamma\in\mathbb{C}$ these are generically non-Hermitean contributions to $H_\kappa$. 
\newline\newline
Along the lines of the interpretation of D-branes from the string worldsheet perspective \cite{Polchinski:1994fq}, the $\sim\tfrac{1}{\kappa}$ behaviour in the exponential of (\ref{nonpert}) might be viewed as adding more boundaries to the disk. As suggested in \cite{anninos2021semiclassical}, if this is the correct interpretation, then the Hartle-Hawking prerscription which assumes a single boundary will have to be non-perturbatively altered. The output of the path integral would instead be a function of multiple boundary lengths, which no longer satisfies the Wheeler-DeWitt equation in any immediate sense. Perhaps all this indicates the necessity for a third-quantization, along with the Pandora's box of questions it carries by its side.

\section{Wheeler-DeWitt \& topology}\label{wdwtop}

The higher-genus contributions $\mathcal{W}^{(h)}(\ell)$ for $h\ge 1$, or more precisely their $\Lambda$-derivative, fail to solve the Wheeler-DeWitt equation (\ref{minisuperspace WdW Lian-Zuckerman}).  In this section, we explore this property. On a higher genus surface,  even upon imposing the conformal gauge, one must path integrate over  over the moduli space, $\mathcal{M}_h$, of the genus-$h$ Riemann surface. Formally, after gauge fixing, one has
\begin{equation}
\mathcal{Z}_\text{grav}[\ell] =  \int_{\mathcal{M}_h} \dd\mu_h \int \mathcal{D}\varphi e^{-S_L[\varphi;\mu_h]} Z_{\text{gh}}[\mu_h] \, \mathcal{O}_{\text{grav}}(\varphi;\mu_h)~,
\end{equation}
where $\mathcal{O}_{\text{grav}}$ is some gravitationally dressed insertion. Here, $\dd\mu_h$ represents the measure for geometric moduli, $\mu_h$, of a genus-$h$ manifold with a single boundary of proper size $\ell$. We have implicitly performed a Laplace transform over the boundary cosmological constant $\mu_B$ to obtain a function of $\ell$. Although one can map this path integral to a matrix model calculation, as done in the previous section,  we are not aware of an efficient way to directly calculate the above gravitational path integral. Nonetheless, there are contributions to the integral whose structure can be anticipated to some degree. Previous related work on this topic includes \cite{Banks:1989df,Moore:1991ir,Ambjorn:2009fm,Anous:2020lka}.

\subsection{A quantum mechanical warm-up}

As a simple warm-up, it is useful to consider a quantum mechanical harmonic oscillator, $x(\tau)$ coupled to a dynamical einbein $\mathfrak{e}(\tau)$. The Euclidean action is given by
\begin{equation}
S_E[x,\mathfrak{e}] = \frac{1}{2} \int \dd\tau \,\mathfrak{e}\left( \mathfrak{e}^{-2} \dot{x}^2 + x^2 \right) + \lambda \int \dd\tau \,\mathfrak{e}~.
\end{equation}
We have allowed for a one-dimensional cosmological constant $\lambda$. We can now consider the path integral of the theory over the Euclidean semi-infinite half-line $\tau \in \mathbb{R}^-$, with boundary conditions that enforce the vanishing of $x$ in the infinite Euclidean past, while satisfying $x(0)=x_0$. Concretely, the path integral reads
\begin{equation}
\mathcal{Z}_{\text{QM}}(x_0) = \frac{1}{\text{vol} \, \mathcal{G}_{\text{diff}}} \int \mathcal{D}\mathfrak{e} \mathcal{D}x  \, e^{-S_E[x,\mathfrak{e}]}~.
\end{equation}
We can consider the problem in a fixed gauge where $\mathfrak{e}=1$, where the path integral over $x$ yields the ground state wavefunction $\Psi_g(x_0)$. Given that we have taken the Euclidean space to be the half-line, there are no geometric moduli left over in the path integral over $\mathfrak{e}$, so taking $\mathfrak{e}=1$ fully fixes the gauge. In order to obtain a finite answer, we must also tune the value of $\lambda$ to cancel against any local contribution that comes from the path integral over $x$. Thus we find that for `trivial' Euclidean topology $\mathcal{Z}_{\text{QM}}(x_0)=\Psi_g(x_0)$ solves the Schr\"odinger equation at vanishing energy. This is the quantum mechanical analogue of the disk path integral satisfying the Hamiltonian constraint.
\newline\newline
Now let us consider instead, the same setup but on a Euclidean interval $\tau \in (-1,0)$ with  $x(-1)=x_i$ and $x(0)=x_f$. In this case, we can only fix the gauge $\mathfrak{e}=T$ up to a modulus $T$ capturing the proper length between the initial and finite time. The Jacobian associated to our gauge fixing choice has been worked out in Chapter 9 of \cite{Polyakov:1987ez}. After the dust settles, one finds the expression 
\begin{equation}
\mathcal{Z}_{\text{QM}}(x_i,x_f;\lambda) = \int_{\mathbb{R}^{+}} {dT} \, e^{-\lambda T} \, G_E(x_i,x_f;T)~,
\end{equation}
where $G_E(x_i,x_f;T)$ is the Euclidean propagator of the quantum harmonic oscillator over a Euclidean time span $T$. Explicitly, 
\begin{equation}
G_E(x_i,x_f;T) = \frac{1}{\sqrt{2\pi \sinh T}} \times \exp\left({-\frac{1}{2}\coth T \left(x_i^2+x_f^2\right) + x_i x_f \text{csch}\, T  }\right)~.  
\end{equation}
Once again, local divergences are absorbed in the bare coupling $\lambda$, but now we are not required to tune its value to obtain a finite result. We note that $G_E(x_i,x_f;T)$ solves the Euclidean time Schr\"odinger equation, viewed as a function of $x_f$ and $T$. However, upon integrating over $T$, the resulting function of $x_f$ is no longer solving a Schr\"odinger equation. The result, instead, is given by the resolvent 
\begin{equation}
\mathcal{Z}_{\text{QM}}(x_i,x_f;\lambda) = \sum_{n\in \mathbb{N}} \frac{\psi^*_{E_n}(x_i)\psi_{E_n}(x_f)}{E_n+\lambda}~,
\end{equation}
where the $E_n = \frac{1}{2}+n$ are the energy levels of the harmonic oscillator, and $\psi_{E_n}(x_f)$ the energy eigenstates. Crucially, the resolvent $\mathcal{Z}_{\text{QM}}(x_i,x_f;\lambda)$ does not satisfy a Schr\"odinger equation, but rather the resolvent equation
%\footnote{It should be noted that we should view $\mathcal{Z}_\lambda(x_i,x_f) $ as an ordinary c-number rather than a vector in a Hilbert space.}
\begin{equation}
\left( \frac{1}{2}\left(-\partial_{x_i}^2 + x_i^2  \right)  + \lambda\right)\mathcal{Z}_{\text{QM}}(x_i,x_f;\lambda) = \delta(x_i-x_f)~.
\end{equation}
 The gravitational path integral involving a geometric modulus yields an object that no longer solves the Schr\"odinger equation. The term that disrupts this is the $\delta(x_i-x_f)$ which is sensitive to the collision of the initial and final positions. The fact that we integrate over positive $T$ only, is the quantum mechanical counterpart to integrating only over positive lapse functions in the gravitational path integral. It is worth noting that if we consider the behaviour of $\mathcal{Z}_{\text{QM}}(x_i,x_f;\lambda)$ as a function of $x_i$ in a region supported away from $x_f$, $\mathcal{Z}_{\text{QM}}(x_i,x_f;\lambda)$ will satisfy a  Schr\"odinger problem with energy $-\lambda$. Moreover, one has the relation
 \begin{equation}
\int_{\mathbb{R}} \dd x_l \mathcal{Z}_{\text{QM}}(x_i,x_l;\lambda) \mathcal{Z}_{\text{QM}}(x_l,x_f;\lambda) = - \partial_\lambda \mathcal{Z}_{\text{QM}}(x_i,x_f;\lambda)~.
 \end{equation}
 The above replaces the standard transitive property of an ordinary transition amplitude $\mathcal{A}(x_i,x_f;T)$ in a quantum mechanical theory.

\subsection{Two-dimensional quantum fields and gravity}
%\textcolor{Red}{Comment on multiple operator insertions and WdW failure.}
The previous warm-up exercise gives us a hint as to where the failure might lie for the higher topology contributions to the disk path integral. For the ordinary disk path integral at genus zero, both the quantum field theoretic and quantum gravity path integrals generate Schr\"odinger wavefunctionals. At higher genus, however, we can no longer fix the entire geometry up to a conformal factor.  We have  to further integrate over geometric moduli as well. In what follows we will consider the path integral for fields on an annulus and a disk with a single handle. 
\newline\newline
{\textbf{Quantum fields.}} Let us imagine first placing a conformal field theory on an annulus whose ratio of boundary $S^1$ sizes is $T\in\mathbb{R}^+$. We impose Dirichlet boundary conditions for the fields at each end. The path integral generates a transition amplitude $\mathcal{A}(\varphi_i(u),\varphi_f(u);T)$ between the two field configurations, $\varphi_i(u)$ and $\varphi_f(u)$. Moreover, as a function of one of the end points and $T$, the transition amplitude $\mathcal{A}(\varphi_i(u),\varphi_f(u);T)$ solves the field theoretic Euclidean time Schr\"odinger problem. As a slightly more elaborate example, we can consider a quantum field theory on a disk with a single handle and an $S^1$ boundary of length $\ell$. Here the path integral produces a functional, $\mathcal{Z}^{(1)}[\varphi(u),\ell;\tau]$, of the Dirichlet data $\varphi(u)$ with $u$ parameterising the boundary $S^1$, and the additional moduli are denoted by $\tau$. We can view $\mathcal{Z}^{(1)}[\varphi(u),\ell;\tau]$ as a wavefunctional evolving with respect to a Hamiltonian that generates the Euclidean time evolution labeled by $\ell$. From the quantum field theoretic perspective, then, the path integral on a disk with a handle generates a state on the spatial $S^1$ Hilbert space, much like it would on the ordinary disk.
\newline\newline
{\textbf{Quantum gravity on the annulus.}} Let us consider first the gravitational path integral, $\mathcal{Z}_{\mathcal{A}_2}[{\ell}_1,{\ell}_2] $, over an annulus with boundaries of modulus $T\in\mathbb{R}^+$. In this case, we have a single geometric modulus, $T \in \mathbb{R}^+$, parameterising the geometry of the annulus.  Once we integrate over the geometric modulus, $T$, however, we no longer generate a solution to the Schr\"odinger problem, but rather a field theoretic counterpart to the quantum mechanical resolvent. As in the quantum mechanical problem, there are end points in the integral over $T$ that disrupt the Schr\"odinger problem. This can be checked explicitly for the case of quantum gravity with $\Lambda>0$ coupled to a minimal model. In this case, the annulus path integral is known explictly. For the case of pure gravity it is given by (\ref{l1l2}), and one can check explicitly that it does not naively satisfy the Wheeler-DeWitt equation (\ref{minisuperspace WdW Lian-Zuckerman}) as a function of $\ell_1$. However, noting the interesting formula from \cite{Moore:1991ir,Ginsparg:1993is}, we can recast the annulus amplitude (\ref{l1l2}) in the following way
\begin{equation}\label{GMformula}
\mathcal{Z}_{\mathcal{A}_2}[\tilde{\ell}_1,\tilde{\ell}_2]   = 2 \Theta(\tilde{\ell}_2-\tilde{\ell}_1) \sum_{n=0}^\infty (-1)^n (2n+1)I_{n+\frac{1}{2}}(2\tilde{\ell}_1) \,K_{n+\frac{1}{2}} (2\tilde{\ell}_2) + ( \tilde{\ell}_1 \leftrightarrow \tilde{\ell}_2)~.
\end{equation}
If we take the limit of $\tilde{\ell}_2 \to 0^+$ while keeping $\tilde{\ell}_1$ fixed, it appears that as a function of $\tilde{\ell}_1$, $\mathcal{Z}_{\mathcal{A}_2}[\tilde{\ell}_1,\tilde{\ell}_2]$ can be expressed as a superposition of an infinite number of permissible Wheeler-DeWitt wavefunctions.
%(Again, $\mathcal{Z}_{\mathcal{A}_2}[\ell_1,\ell_2]$ should be viewed as a c-number.) 
However, this is not quite correct since the required Bessel-$K$ indices $\nu_t$ appearing in (\ref{GMformula}) span over all half-integers. In contrast, we saw in section \ref{LZsec} that according to the BRST analysis of Lian-Zuckerman, the index $\nu$ of the modified Bessel function, $K_\nu(z)$, is only permitted to take the values $\tfrac{1}{2}+n$ with $n\in\mathbb{N}$ and $n \neq  1 \,\text{mod} \, 3$. 
\newline\newline
So, (\ref{GMformula}) requires additional values of $\nu$ that are not found in the Lian-Zuckerman BRST cohomology. In fact, taking these values of $\nu$ and combining (\ref{minisuperspace WdW Lian-Zuckerman}) with (\ref{con dim primaries}) would yield non-integer weights for the Liouville dressing operators making it impossible to combine them with ghost-fields and derivatives of the Liouville field in such a way that the net conformal weight is one. On the other hand, as noted in \cite{Martinec:1991ht}, the corresponding boundary weights (\ref{conformal weight boundary operator}), at least in the case of pure two-dimensional quantum gravity with $\Lambda>0$, are integer valued. It was consequently argued \cite{Martinec:1991ht}, that the remaining values of $\nu$ appearing in (\ref{GMformula}), with $n = 1 \,\text{mod} \, 3$, stem from boundary operators appropriately dressed with $\mathfrak{bc}$-ghost fields. The simplest example of a boundary operator is the proper length $\ell$ of the boundary itself, and its associated coupling is $\mu_B$ corresponding to the value $\nu=\frac{3}{2}$. Others are reported in \cite{Kostov:2003cy}.
\newline\newline
{\textbf{Quantum gravity on a higher-genus disk.}} We now move on to the genus-one disk gravitational path integral. We will not be able to analyse this example in any detail. Instead, we will consider particular contributions to the gravitational path integral from certain limiting regions in the moduli space. One particular contribution comes from the regime where there is a long thin tube connecting a torus with a small disk excised to the disk with another small disk excised. We can view this \cite{Cohen:1986pv,Moore:1986rh} as connecting a punctured disk to a punctured torus through the propagation of on-shell states. As noted in section \ref{largelsec}, if we consider the leading  behaviour in the large-$\ell$ limit for each genus, the result bears strong similarity to the result in topological gravity detailed in appendix \ref{topgrav}. In this latter case, the behaviour is entirely topological and determined by the one-point genus-$h$ intersection numbers $\langle\tau_{3h-2}\rangle_h = \left(\tfrac{1}{24}\right)^h \frac{1}{h!}$ uncovered in  \cite{witten1993two}. Indeed, expanding (\ref{Wtop}) in a small-$\kappa$ expansion we have
\begin{equation}\label{Whtop}
\mathcal{W}^{(h)}(\tilde{\ell}) = \frac{e^{-\tilde{\ell}/2}}{\sqrt{2\pi}}  \,\tilde{\ell}^{3h-\frac{3}{2}} \left(\frac{\tilde{\kappa}}{2}\right)^{2h-1}\times \left( \frac{1}{24}\right)^h\frac{1}{h!}~.
\end{equation}
The $\tilde{\ell}$-dependent pre-factor of the above expression accounts for the fact that we are dealing with a Riemann surface with a boundary. However,  this does not modify the topological content of $\langle\tau_{3h-2}\rangle_h$ which also appears in (\ref{Whtop}).
\newline\newline
{\textbf{Comment on the lapse contour.}} As a final remark, the failure of $\mathcal{W}^{(h)}(\ell)$ to solve the Wheeler-DeWitt equation (\ref{minisuperspace WdW Lian-Zuckerman}) indicates the gravitational path integral is not integrating the lapse function along the imaginary axis. Had this been the case, it would have implemented the Hamiltonian constraint as a $\delta$-function. It is interesting to ask whether there is any computation in the matrix model that implements such a contour for the lapse-function. It would seem that such an operation would have project onto trivial topology. From the perspective of the planar expansion, we would like to isolate the $\sim\kappa^{-1}$ term. To do so we can perform a contour integral around the origin of $\kappa$, namely $\oint d\kappa \,\mathcal{W}_\kappa(\ell)$. To do so, we need an understanding of the full topological expansion $\mathcal{W}_\kappa(\ell)$ in the complex-$\kappa$ plane at least for a neighbourhood near the origin. However, the topological expansion for pure two-dimensional quantum gravity with $\Lambda>0$ is not Borel summable. An alternative perspective is that we should not require our state to satisfy the Hamiltonian constraint, but rather define physical states as equivalence classes related by a shift proportional to the Hamiltonian constraint \cite{Higuchi:1991tm}. Again, it would be interesting to understand how such a group averaging procedure is encoded in the matrix model.

\section{Timelike Liouville theory, a comparison}\label{outlook}

Rather than an outlook, we would like to end our discussion by briefly comparing our discussion to the  case of a unitary two-dimensional conformal field theory with large and positive central charge, $c$, coupled to $\Lambda>0$ quantum gravity. In the conformal gauge this theory can be mapped to a timelike Liouville theory. A cosmological perspective for this theory was offered in \cite{Anninos:2024iwf,Anninos:2025fer}.\footnote{A conformal bootstrap approach to timelike Liouville theory has been undertaken in \cite{Kostov:2005kk,Schomerus:2003vv,Zamolodchikov:2005fy,Ribault:2015sxa,Collier:2023cyw,Giribet:2026gao}. An approach based on the path integral has been offered in \cite{Anninos:2021ene,Muhlmann:2022duj}, with a recent more rigorous treatment based on probabilistic methods offered in \cite{Chatterjee:2025yzo,Chatterjee:2026zmb}.} An important distinction from the case of pure two-dimensional gravity with $\Lambda>0$ studied throughout the main text is the existence of a semiclassical limit yielding a two-dimensional de Sitter spacetime vacuum plus a family of big bang/crunch cosmologies \cite{Anninos:2024iwf}. The large-$c$ regime controls the size of the gravitational fluctuations.
\newline\newline
{\subsection{Wheeler-DeWitt}} For the large-$c$ theory, it was found in \cite{Anninos:2024iwf} that the solutions to the Wheeler-DeWitt equation display oscillatory behavior at large $\ell$, rather than the exponentially decaying behaviour displayed by  (\ref{genus zero path integral heat kernel}). Furthermore, in contrast to the case at hand, for sufficiently small matter excitations, there exist Wheeler-DeWitt solutions in the large-$c$ theory which vanish at small $\ell$. These are closer in spirit to the structure envisioned by Hartle and Hawking \cite{hartle1983wave}. In addition, there are other solutions to the Wheeler-DeWitt equation that diverge at small $\ell$, but these also oscillate at large $\ell$. We view the oscillatory behaviour at large $\ell$ for the timelike Liouville case as being more characteristic of a cosmological WKB type behavior. Moreover, a candidate inner product for the Wheeler-DeWitt Hilbert space has been considered in \cite{Anninos:2025fer,Anninos:2026hia}, whereby one pairs two wavefunctions evaluated at a fixed trace of the extrinsic curvature, $K$ (one can view $K$ as the York time variable \cite{york1972role}). An interesting property of this pairing is that the dependence on $K$ can drop out of the norm for the cases studied so far. This will not occur for the case of pure two dimensional gravity with $\Lambda>0$. Let us unpack this. Given that $K$ and $\ell$ are conjugate variables, we have $K\propto\mu_B$. Going to the fixed $\mu_B$ representation amounts to a Laplace transform on $\ell$, with measure $\tfrac{d\ell}{\ell}$. Performing this for the wavefunctions (\ref{BesselK WF}) in the $\Lambda=1$ frame yields
\begin{equation}\label{psimu}
%\Psi_\alpha(\tilde{\mu}_B) = -\frac{\pi   \csc (\pi  \nu_\alpha )}{2 \sqrt{\tilde{\mu}_B^2-4} }{\left(2^{\nu_\alpha } \left(\sqrt{\tilde{\mu}_B^2-4}+\tilde{\mu}_B \right)^{-\nu_\alpha }-2^{-\nu_\alpha } \left(\sqrt{\tilde{\mu}_B^2-4}+\tilde{\mu}_B \right)^{\nu_\alpha }\right)}~.
\Psi_\alpha(\tilde{\mu}_B) = -\frac{\pi   \csc (\pi  \nu_\alpha )}{2 \nu_\alpha }{\left(2^{\nu_\alpha } \left(\sqrt{\tilde{\mu}_B^2-4}+\tilde{\mu}_B \right)^{-\nu_\alpha }+2^{-\nu_\alpha } \left(\sqrt{\tilde{\mu}_B^2-4}+\tilde{\mu}_B \right)^{\nu_\alpha }\right)}~.
\end{equation}
We note the manifest reflection symmetry, $\nu_\alpha \leftrightarrow -\nu_\alpha$, in the above expression, and we have dropped the overall normalisation. To compute the Laplace transform, we have assumed $\tilde{\mu}_B>2$ and regulated the small $\ell$ divergence by analytic continuation. It is clear from (\ref{psimu}), that the pairing implemented in \cite{Anninos:2025fer,Anninos:2026hia}, namely $\mathcal{N}_\alpha \equiv \Psi_\alpha(\tilde{\mu}_B)\Psi^\dag_\alpha(-\tilde{\mu}_B)$,  will not yield a $\tilde{\mu}_B$ independent result for non-vanishing $\nu_\alpha$. For this to be the case, we would need to somehow break the reflection symmetry $\nu_\alpha \leftrightarrow -\nu_\alpha$, which is known not to happen in spacelike Liouville theory. Depending on how we treat the theory, such violations of reflection symmetry can occur for the timelike Liouville case (for recent discussions see \cite{Anninos:2024iwf,Anninos:2025fer,Chatterjee:2026zmb,Giribet:2026gao}). 
\newline\newline
There is currently little understanding of the large-$c$ theory on a higher genus manifold. The main obstruction is that a naive integration over the genus-$h$ moduli space will lead to strong ultraviolet divergences that cannot be treated with standard local methods \cite{Anninos:2022ujl}. The origin of these divergences stems from the conformal field theory matter fields that get highly excited along small cycles that are perceived as an effective high temperature regime. It would be very interesting to see whether taming such divergences, perhaps through a modification of the moduli space integration contour, will lead to a Hartle-Hawking type wavefunction that also offends the Wheeler-DeWitt equation.

{\subsection{Sphere path integral}} The sphere path integral of timelike Liouville theory was computed from a path integral approach in \cite{Anninos:2021ene,Muhlmann:2022duj}. Using the contour prescription of \cite{gibbons1978path,polchinski1989phase}, it is found that the sphere path integral now takes the form \cite{Anninos:2021ene}
\begin{equation}\label{Ztlt}
\mathcal{Z}_{\text{grav}}[\Lambda] = \mathcal{N}_0 \, e^{2\vartheta} \Lambda^{-\frac{q}{\beta}}~, 
\end{equation}
The prefactor $\mathcal{N}_0$ can be computed systematically in the loop expansion, upon selecting a regularization scheme. Here we have $q\equiv\beta^{-1}-\beta$, and we further have the relation
\begin{equation}
\frac{q}{\beta} = \frac{1}{12}\left(c+\sqrt{(c-25)(c-1)}-25 \right) \approx \frac{c}{6} - \frac{19}{6} + \ldots~.
\end{equation}
 In the second part, we have expanded at large $c$, and the first term can be recognised as the matter entanglement entropy. Interestingly, an inverse Laplace transform of (\ref{Ztlt}) to the fixed volume, $\upsilon$, ensemble yields
\begin{equation}\label{Ztlt}
\mathcal{Z}_{\text{grav}}[\upsilon] = \frac{\mathcal{N}_0}{\Gamma\left({\beta}^{-2}-1\right)} \, e^{2\vartheta}  \, \upsilon^{\frac{q}{\beta}-1}~. 
\end{equation}
We further note that in the fixed volume ensemble, the large $c$ expansion yields
\begin{equation}\label{Ztlt2}
\log \mathcal{Z}_{\text{grav}}[\upsilon] \approx 2\vartheta + \left( \frac{c}{6} + \frac{1-26}{6} + \ldots \right)  \log \upsilon + \text{const}
%\frac{\mathcal{N}_0}{\Gamma\left(\frac{Q}{b}\right)} \, e^{2\vartheta}  \, \upsilon^{-1+\frac{Q}{b}}~. 
\end{equation}
and perhaps we can interpret the $\tfrac{1-26}{6}$ as contribution to the entanglement from the  Liouville and $\mathfrak{bc}$-ghost sectors \cite{Anninos:2021ene}. If we are to demand that the fixed area partition function $\mathcal{Z}_{\text{grav}}[\upsilon]$ is positive, as it was for pure two dimensional gravity, then we can fix that $\mathcal{N}_0>0$ yielding a positive $\mathcal{Z}_{\text{grav}}[\Lambda]$  (see also \cite{Blanco:2026zhq}). 
\newline\newline
A notable difference is that the Laplace transform of (\ref{Ztlt}) back to the fixed $\Lambda$ ensemble is convergent in this case, since there is no small $\upsilon$ divergence.

\subsection{Liouville line defects} 

Finally, it is worth noting that spacelike Liouville theory admits interesting line defects recently analysed in \cite{Abdalla:2026wdx}. These are given by integrating the operator $\mathcal{O}_{b/2}$ over some curve $\mathcal{C}$ inside the spacetime. Notice that this is different from the case of a manifold with a boundary. One can analyse the properties of these line defects in the semiclassical regime, and they lead to a small localised jump in the extrinsic curvature governed by the coupling to the line defect. From a gravity perspective it might be natural to dress them with matter fields or $\mathfrak{bc}$-ghosts. Interestingly, in the semiclassical $b\to0^+$ limit, one can use these line defects to create a closed Euclidean world consisting of two hyperbolic disks whose boundary intersects at the line defect \cite{Abdalla:2026wdx}. Perhaps these line defects, decorated with the matter fields, can play the role of a worldline observatory \cite{Anninos:2011af} measuring the ambient space of the closed world. 
\newline\newline
It is of particular interest is to extend the analyses of \cite{Apresyan:2018pvl,Sarkissian:2009aa,Abdalla:2026wdx} to timelike Liouville theory, at least in the semiclassical regime.\footnote{Perhaps one should consider even more general line defects which combine the Liouville and matter sectors in a conformally invariant way. Moreover, it might also be worth considering multiple defects.} For the timelike Liouville case, there is already a semiclassical sphere saddle \cite{Anninos:2021ene} in the absence of the line defect and the insertion of the defect will create a small cusp, corresponding to a jump in the normal derivative of the Liouville field. From the action formalism perspective, adding a line operator $\mathcal{O}_\nu = \nu\int \dd\xi \sqrt{\tilde{h}} e^{\alpha \varphi}$ to the timelike Liouville action (see (2.6) of \cite{Anninos:2021ene} for conventions) yields the equation of motion
\begin{equation}\label{tltD}
    \frac{1}{2\pi} \tilde{\nabla}^2 \varphi  - \frac{1}{4\pi} q \tilde{R}  +2 \beta \Lambda e^{2\beta\varphi} +  \alpha \nu \cosh\vartheta_0 \delta(\vartheta-\vartheta_0)  e^{\alpha\varphi} = 0~.
\end{equation}
For the saddle point approximation to be valid, we require that $\Lambda$ scales as $\beta^{-2}$, and $\alpha\nu$ scales as $\beta^{-1}$. To leading order we can further take $q\approx\beta^{-1}$. For our fiducial metric $\tilde{g}_{ij}$ we have taken
\begin{equation}
\dd \tilde{s}^2 = \frac{\dd\vartheta^2 + \dd\xi^2}{\cosh^2\vartheta}~, \quad\quad \vartheta \in \mathbb{R}~, \quad\quad \xi \sim \xi+2\pi~,
%\dd\vartheta^2 + \sin^2\vartheta \dd\phi^2~, \quad\quad \vartheta \in (0,\pi)~, \quad\quad \phi \sim \phi+2\pi~.
\end{equation}
such that the physical metric is given by $e^{2\beta\varphi}\tilde{g}_{ij}$. The line defect resides along the $\vartheta=\vartheta_0$ closed curve. Away from the delta-function, the solution to (\ref{tltD}) gives a physical metric which describes a portion of the round two-sphere of scalar curvature $R={8\pi\beta q^{-1}\Lambda}$. The jump in the normal derivative is $\partial_\vartheta\varphi|_{\vartheta_0-\varepsilon}^{\vartheta_0+\varepsilon} = -2\pi \alpha \nu \text{sech}\vartheta_0 e^{\alpha \varphi(\vartheta_0,\xi)}$. Taking the semiclassical value $q\approx\beta^{-1}$, and assuming a $\xi$-independent solution, we find
\begin{equation}\label{defect}
 e^{2\beta\varphi_\pm(\vartheta,\xi)} =  \frac{1}{4\pi\Lambda \beta^2}  \frac{\cosh^2 \vartheta}{\cosh^2 (\vartheta-\vartheta_0\pm\delta)}~.
%\quad\quad\quad e^{2\beta\varphi_-} = \frac{1}{4\pi\Lambda \beta^2}  \left(\frac{\cosh x}{\cosh (x-x_0-\delta)} \right)^2~,
\end{equation}
where the $\pm$ label indicates whether we are above or below $\vartheta_0$. The parameter $\delta$ is fixed by the jump condition, and reads
\begin{equation}
\tanh \delta = \alpha\nu \beta \pi  \,\text{sech}\vartheta_0 \left(\frac{\cosh \vartheta_0}{\cosh \delta} \right)^{\frac{\alpha}{\beta}} \left( \frac{1}{4\pi\Lambda \beta^2} \right)^{\frac{\alpha}{2\beta}} \, ~.
%e^{\alpha \varphi_0}~.
\end{equation}
We note that $\delta=0$ for $\nu=0$. The volume of the deformed physical two-sphere saddle is now $\text{vol} \,S^2= (\Lambda\beta^2)^{-1} (1-\tanh\delta)$, such that a $\nu>0$ deformation of `positive energy' slightly reduces the volume. We further note that the total scaling dimension of the integrated cosmological line operator $\mathcal{O}_\beta$ is $\Delta^{(\text{tot})}_\beta = 1-\tfrac{\beta^2}{2}$ which is slightly relevant. In contrast, the analogous line operator in spacelike Liouville is slightly irrelevant. Of course, $\alpha$ can be adjusted to render the defect marginal also. If we instead have two line defects, or more, they will interact with each other, and maybe this can be viewed as a gravitational exchange from the underlying gravitational perspective. 
\newline\newline
It is also interesting to note that in spacelike Liouville there are topological  line defects \cite{Apresyan:2018pvl} separating two spacelike Liouville theories endowed with different values of the cosmological constant $\Lambda$. Again, we see no immediate obstruction to the construction of such  line operators in timelike Liouville.\footnote{As before, we might imagine dressing these with matter theories that admit topological line defects, such as the Schwinger model with charge-$p$ fermions recently studied on a de Sitter background \cite{Aguilera-Damia:2026dbk}. Since the total stress-tensor vanishes  in  theories of gravity on compact spatial slices, there is a naive sense  that any line-defect is topological, or better stated that the gravitationally dressed displacement operator is BRST exact. However, one should still distinguish those originating from topological line defects in the underlying matter theory, from those that are rendered topological upon gravitational dressing.} Following the arguments of section 4 in \cite{Apresyan:2018pvl}, one  finds that such cosmological constant changing defects are  topological for timelike Liouville theory. Physical quantities will now also depend on the dimensionless ratio $\tfrac{\Lambda_1}{\Lambda_2}$ of the two cosmological constants. 
\newline\newline
Perhaps more interestingly, along the lines of Coleman and De Luccia's instanton \cite{Coleman:1980aw} in the thin-wall limit, would be a gravitationally dressed line defect that keeps the induced physical metric fixed across the interface.\footnote{Going slightly away from the thin-wall limit will require a richer description, that is captured by additional effective field theory terms in the defect Lagrangian. This is related, but not equivalent, to the discussion in appendix \ref{EFTb}.} This would correspond to a generalisation of the line defects in \cite{Abdalla:2026wdx} for timelike Liouville theory, but now connecting two theories with different values of $\Lambda$. (The defect (\ref{defect}) would correspond to the case where the wall connects two vacua with equal $\Lambda$.) Semiclassically, the jump in the physical extrinsic curvature will be akin to the Israel junction condition \cite{Israel:1966rt} setting the domain/bubble wall tension. This allows us to analyse the Coleman-De Luccia instanton in an ultraviolet finite framework, using the language of line defects and conformal bootstrap methods. The Coleman-De Luccia bounce action might then be viewed as a type of Affleck-Ludwig entropy \cite{Affleck:1991tk}.
\newline\newline
Bearing all this in mind, then, the theoretical laboratory of timelike Liouville equipped with such defects can aid us in sharpening puzzles  associated to the role of observers, and the global structure of eternally inflating Universes with multiple vacua \cite{Guth:2007ng}. These features will be explored in future work.

%In fact, the line defect of \cite{Apresyan:2018pvl} can also be treated semiclassically. 

\section*{Acknowledgements}

We are very grateful for discussions with Tarek Anous, Victor Gorbenko, Chris Herzog, Joel Karlsson, Edward Mazenc, Olga Papadoulaki, and Edgar Shaghoulian, and ChatGPT. We are particularly indebted to Beatrix M\"uhlmann for many illuminating discussions on these issues. And also Panagiotis Betzios for initial collaboration and insightful comments. We acknowledge fruitful discussions during the workshop ``Observers, wormholes and complex saddles in cosmology", organized at the Bernoulli Center for Fundamental Studies (EPFL, Lausanne) from 18--22 May 2026. D.A. is funded by the Royal Society under the grant “Concrete Calculables in Quantum de Sitter,” the STFC consolidated grant ST/X000753/1, and the KU Leuven grant ``Holography on the horizon" C16/25/010. S.B. is funded by STFC under the grant reference STFC/2928477.

\appendix

\section{A simple effective field theory with a boundary} \label{EFTb}

In this appendix, we study the effective field theory for a quantum field theory on a spacetime with a boundary (see \cite{Deutsch:1978sc,Diehl:1981jgg} for early work). For the sake of simplicity, we will consider a shift symmetric scalar field $\varphi$ in two spacetime dimensions on a manifold with a boundary. We will also assume that the theory respects the $\mathbb{Z}_2$ symmetry $\varphi \to -\varphi$.

\subsection{Bulk effective field theory}

To begin with, let us consider the case of the theory on $\mathbb{R}^2$, such that the spatial slice is the full real line $\mathbb{R}$. The Lagrangian is organised in terms of a derivative expansion
\begin{equation}\label{beft}
S_{\text{EFT}}  = \int_{\mathbb{R}^2} \dd^2x \left( \frac{1}{2} \partial_\mu \varphi\partial^\mu \varphi  + \sum_{n=2}^\infty \gamma_n \left( \partial_\mu \varphi\partial^\mu \varphi\right)^n  + \ldots  \right)~.
\end{equation}
The $\gamma_n$ are real-valued couplings for increasingly irrelevant operators. The on-shell condition is $\square \varphi = 0$. Local field redefinitions are given by
\begin{equation}\label{frdef}
\varphi \to \varphi + \varepsilon \, \mathcal{F}(\partial_\mu \varphi,\ldots)~,%\textbf{}
\end{equation}
such that the variation in the action, to leading order in the derivative expansion, goes as
\begin{equation}
\delta S_{\text{EFT}} = -\varepsilon \int_{\mathbb{R}^2} \dd^2x \, \mathcal{F} \, \square \varphi  + \ldots
\end{equation}
Due to the above, we can remove any coupling constants in $S_{\text{EFT}}$ that vanish on-shell. In particular,  this implies that operators of the type $\{\square \varphi \square \varphi, \square \partial_\mu\varphi \square  \partial^\mu \varphi,\ldots \}$ can be removed with local field redefinitions, and are hence we do not include them in $S_{\text{EFT}}$. 
%One could also consider terms with distributed derivatives, such as $\partial_\mu\partial_\nu\varphi \partial^\mu\partial^\nu\varphi$, however, upon integrating by parts these can also be put into the general form of $S_{\text{EFT}}$. 
%We see that the effective field theory of a shift symmetric scalar in two-dimensions is captured by the list of couplings $\gamma_n$, with $n\in\mathbb{N}$. 
\newline\newline
To complement the Lagrangian perspective, one can calculate some simple $S$-matrix elements, as these are field redefinition invariant \cite{lehmann1955formulierung}. At tree-level, it is clear that any coupling proportional to the leading order equations of motion will not contribute to the scattering amplitude due to the fact that the external states are on-shell. At one loop, there may be scheme dependendent regularization choices but these will only affect local contributions to the $S$-matrix, the non-local remainder again depends only on field-redefinition invariant information.

\subsection{Boundary effective field theory}

We now consider the same theory, but on the half-space $\mathbb{R} \times \mathbb{R}^+$ rather than $\mathbb{R}^2$. In this case, caution must be exercised when performing field redefinitions and integrating by parts as this can yield additional boundary couplings that we previously did not need to worry about. To set up the effective field theory in this case, we include both bulk and boundary contributions with the most general terms. Let us work this out to first non-trival order. The bulk effective field theory to this order reads 
\begin{multline}
S^{\text{bulk}}_{\text{EFT}}  = \int_{\mathbb{R} \times \mathbb{R}^+} \dd^2x  \left( \frac{1}{2} \partial_\mu \varphi\partial^\mu \varphi  + \alpha_1 \left( \partial_\mu \varphi\partial^\mu \varphi\right)^2 + \alpha_2  \partial^\mu \varphi \square \partial_\mu \varphi + \alpha_3  \square \varphi \square \varphi + \right. \\ \left.  \alpha_4 \partial_\mu\partial_\nu \varphi \partial^\mu \partial^\nu \varphi   +\ldots \right) ~.
%+ \alpha_5 \varphi \square\square \varphi
\end{multline}
We have not imposed any on-shell conditions or performed any integration by parts. Prior to imposing any boundary conditions, the most general boundary effective field theory to leading order is
\begin{equation}\label{Sbdygen}
S^{\text{bdy}}_{\text{EFT}}  = \int_{x=0} \dd u \left(  a_1  (\partial_u \varphi)^2  + a_2 (\partial_n \varphi)^2 +   a_3 \partial_u \varphi \partial_n \varphi   +  b_1 \partial^2_u \varphi \partial_n \varphi +  b_2 \partial^2_n \varphi \partial_n \varphi  +  b_3 \partial^2_n \varphi \partial_u \varphi  + \ldots  \right)~,
%a_1 \partial_n \varphi+ a_2 \partial_n^2 \varphi  + a_4 \varphi \partial_n^2 \varphi
\end{equation}
where $\partial_n$ delineates a derivative outwardly normal to the boundary, whilst $\partial_u$ delineates a derivative tangential to the boundary. The boundary coordinate is denoted by $u \in \mathbb{R}$. We note that there are intrinsic and extrinsic boundary couplings, and we have removed terms that are redundant upon integrating by parts along the boundary direction (as the boundary has no boundary). We have also enforced both the $\mathbb{Z}_2$ and shift symmetry of $\varphi$ at the boundary. 
\newline\newline
So far, however, we have not implemented any integration by parts, on-shell conditions, or local field redefinitions on the effective field theory. We have also not specified any boundary conditions on the fields. A local field redefinition of the type (\ref{frdef}) will now transform the bulk action, to leading order in the EFT expansion, as
\begin{equation}
\delta S^{\text{bulk}}_{\text{EFT}}  = \varepsilon \int_{\mathbb{R} \times \mathbb{R}^+} \dd^2x  \partial_\mu \varphi \partial^\mu \mathcal{F} = - \varepsilon \int_{\mathbb{R} \times \mathbb{R}^+} \dd^2x  \mathcal{F} \, \square \varphi + \varepsilon \int_{x=0} \dd u \, \mathcal{F} \partial_n \varphi  ~.
\end{equation}
We see that the boundary coupling $a_3$ can be shifted alongside bulk couplings proportional to the equations of motion. For instance if we take $\mathcal{F}(\varphi)=\square \varphi$, the field redefinition will shift $\alpha_3 \to \alpha_3 - \varepsilon$ along with $b_1 \to b_1 + \varepsilon$ and $b_2\to b_2 + \varepsilon$. So we can use bulk field redefinitions to remove couplings proportional to the equations of motion, at the price of reshuffling boundary couplings. In addition, we can implement boundary local field redefinitions 
\begin{equation}
\varphi |_{x=0}\to \varphi + \varepsilon_b \mathcal{F}_b(\partial_u\varphi,\partial_n\varphi,\ldots)|_{x=0}~.
\end{equation}
The boundary local field redefinitions   have the effect of reshuffling intrinsic boundary couplings. More generally, one can also consider boundary field redefinitions that allow normal as well as tangential derivatives of $\varphi$. Normal derivatives are tied to boundary couplings upon integrating bulk terms by parts. 

\subsection{Simple boundary observables} 

As for the bulk effective field theory, it is useful to consider an observable in the boundary effective field theory. One observable we can imagine is a scattering amplitude on the half-space. As a simple example, let us consider the  following Lorentzian action
\begin{equation}
S_{\text{EFT}} =  \int_{\mathbb{R} \times \mathbb{R}^+} \dd t \dd x \left( -\frac{1}{2} \partial_\mu \varphi\partial^\mu \varphi  + \alpha \left( \partial_\mu \varphi\partial^\mu \varphi\right)^2 \right) +  \gamma\int_{x=0} \dd t \left( \partial_t{\varphi} \right)^2~,
\end{equation}
Consider first the tree-level two-to-two scattering amplitude $\mathcal{S}_{2,2}$. One contribution stems entirely from a bulk interaction and the other from a boundary interaction. The bulk scattering process can occur before or after one or both of the waves reflect off the boundary. The result is a simple application of the method of images, and is given by a sum over various images of the bulk amplitude $\mathcal{S}^{(\text{bulk})}_{2,2}$ for the theory without boundary. In addition to the bulk scattering process, we can have a scattering process localised at the boundary. The most isolated of these is the tree-level one-to-one scattering amplitude $\mathcal{S}^{(\text{bdy})}_{1,1}$. 
\newline\newline
Along with the bulk equations of motion, we also have the modified boundary condition
\begin{equation}
\partial_x \varphi +  2 \gamma\partial^2_t\varphi|_{x=0} = 0~.
\end{equation}
It follows that the tree-level scattering process, $\mathcal{S}_{1,1}$, is given by a non-local reflection phase
\begin{equation}
\mathcal{S}_{1,1} = 2\pi \left( \frac{1+2 i \gamma  k}{1-2 i \gamma  k}\right)\delta(k-k')~,
\end{equation}
where $k>0$ is the absolute value of the spatial momentum for the incoming wave, and $k'>0$ is the absolute value of the outgoing wave. At large spatial momentum, $k>0$, the phase tends to $-1$ which is the Dirichlet problem, while at small momenta it tends to $1$ which is the Neumann problem. Beyond the tree-level, $\alpha$ will also contribute to $\mathcal{S}_{1,1}$. 
\newline\newline
{\textbf{Boundary field redefinition invariance.}} We also note that an intrinsic boundary field redefinition will not affect the non-local structure of the scattering phase. Consider for instance the  boundary field redefinition
\begin{equation}
\varphi|_{x=0} \to \varphi+ \varepsilon_b (\partial_t \varphi)^2|_{x=0}~.
\end{equation}
To understand the modification of the theory, it is convenient to split the bulk field into a bulk piece $\varphi$ and a boundary piece $\phi(t)\equiv\varphi(t,x)|_{x=0}$. Prior to any field redefinitions, the original theory (with $\alpha=0$ for simplicity) can be reformulated to the following
\begin{equation}\label{Stot}
S_0 = -\frac{1}{2} \int_{\mathbb{R}\times\mathbb{R}^+
} \dd t \dd x  \partial_\mu \varphi \partial^\mu \varphi + \int_{x=0} \dd t \left( \gamma (\partial_t{\phi}(t))^2 + i \lambda(t)(\phi(t)-\varphi(t,0) \right)~,
\end{equation}
where $\lambda(t) \in \mathbb{R}$ is a Lagrange multiplier that sets $\phi=\varphi|_{x=0}$. The boundary condition is now formulated in terms of the boundary equations of motion
\begin{equation}\label{beom}
\varphi|_{x=0}=\phi~, \quad\quad \partial_x \varphi|_{x=0} + i \lambda=0~, \quad\quad 2\gamma\partial_t^2 \phi = i \lambda~.
\end{equation}
In this setup, a local boundary field redefinition is given by
\begin{equation}
\phi \to \phi + \varepsilon_b \mathcal{F}_b(\partial_t \phi)~, \quad\quad \lambda \to \lambda~, \quad\quad \varphi \to \varphi~.
\end{equation}
This modifies the total action $S_0$ in (\ref{Stot}) by a term proportional to the boundary equations of motion (\ref{beom}) up to total derivatives in $t$. In this sense, on-shell data, such as the boundary scattering matrix, will be insensitive to field redefinitions. Moreover, there will exist  couplings such as $\gamma$ that cannot be removed with boundary field redefinitions, and many others that do not source terms proportional to the boundary equations of motion.

\section{Details for genus-one and higher} \label{Heat kernel calc}

In this appendix, we present the calculations for the genus-one (\ref{The wavefunction genus one}) and genus-two wavefunction (\ref{Genus two wavefunction}), obtained via a semiclassical analysis of the path integral (\ref{path integral heat kernel}).  We also present the results for some higher-genus wavefunctions using the heat kernel method.

\subsection{Genus-one and -two using path integral method}\label{appPI}

\textbf{Classical trajectory.} Firstly, we must  correct the classical path. To the order we are interested in, we have
\begin{equation}\label{classical path genus two}
    z_{\text{cl}}(\tau)  \approx z_0 - \frac{1}{8\sqrt{z_0}}\left(\ell - \tau\right)\tau\kappa ^2 -\frac{  \tau (\ell-\tau) \left( z_0 \left(\ell^2+\ell \tau-\tau^2\right)+16\right)}{768 z_0^3}\kappa ^4+\ldots~.
    %z_{\text{cl}}(\tau)  \approx z_0 - \frac{2}{\sqrt{z_0}}\left(\ell - \tau\right)\tau\kappa ^2 -\frac{  \tau \left(  \ell-\tau+\left(\ell^3-2\ell \tau^2+\tau^3\right)z_0\right)}{3 z_0^3}\kappa ^4+\ldots~.
\end{equation}
The corresponding on-shell Euclidean action is given by
\begin{equation}
S_{\text{cl}} = \frac{\ell\sqrt{z_0}}{2}-\ell\left(\frac{1}{48z_0^2}+\frac{\ell^2}{384 z_0}\right)\kappa^2-\ell\left(\frac{49}{2304 z_0^{\frac{9}{2}}}+\frac{\ell^2}{1152 z_0^{\frac{7}{2}}}+\frac{\ell^4}{30720 z_0^{\frac{5}{2}}}\right)\kappa^4+\ldots~.
\end{equation}
As a crosscheck for the above on-shell action, we can compute the corresponding classical path for more general end points $z_1$ and $z_2$, and show it satisfies both the Hamilton-Jacobi equation at the correct order in $\kappa$.% and recovers the diagonal boundary conditions of interest under the coincident point limit. 
%\subsection{One-loop determinant}
\newline\newline
\textbf{One-loop determinant.} We must also compute the corrected one-loop functional determinant for quadratic fluctuations about the classical path (\ref{classical path genus two}). This can be computed using the method of Gelfand and Yaglom \cite{gel1960integration}. To the order we are interested in, one finds
\begin{equation}
    \text{det}^{-\frac{1}{2}} {\mathcal{O}_\kappa} \approx \ell^{-\frac{1}{2}}+\frac{ \ell^{\frac{3}{2}}}{96 z_0^{\frac{3}{2}}}\kappa ^2+\frac{\ell^{\frac{3}{2}} \left(64+3 \ell^2 z_0\right)}{6144 z_0^4}\kappa ^4 + \ldots~. %\mathcal{O}(\kappa ^6)
\end{equation}
It is straightforward to use the Gelfand-Yaglom method to compute the determinant to higher orders.
\newline\newline
\textbf{Two-loop correction.} At genus-two and higher we must also include contributions from higher loop Feynman diagrams. At genus two, the only such contribution comes from a quartic vertex  yielding the (disconnected) figure of eight diagram. Upon expanding the action in the small fluctuation $\zeta(\tau)$ one finds the quartic interaction term
\begin{equation}
S_{\text{int}} =     -\frac{5\kappa^4}{256 z_0^{\frac{7}{2}}}\int_0^\ell \text{d}\tau \, \zeta(\tau)^4~.
\end{equation}
There is also a $\sim\kappa^3$ cubic vertex but this would only contribute to higher order in $\kappa$. The propagator to leading order in $\kappa$ is given by
\begin{equation}\label{propagator}
G(\tau,\tau') = 
\begin{cases}
-\frac{\tau(\ell-\tau' )}{\ell}+\mathcal{O}(\kappa^2)~, \quad\quad &\tau<\tau'~, \\
-\frac{\tau'(\ell-\tau)}{\ell}+\mathcal{O}(\kappa^2)~, & \tau>\tau'~.
\end{cases}
\end{equation}
From these we obtain the two-loop contribution
\begin{equation}\label{quartic vertex leading order}
\mathcal{I}_2 = \frac{15\kappa^4}{256 z_0^{\frac{7}{2}}} \int_0^\ell \text{d}\tau \,G(\tau,\tau)^2~. 
%&=-\frac{l^3\,\kappa^3}{12288 z^{\frac{7}{2}}}\,,
\end{equation}
%({\color{blue}what about connected corrections to the propagator?})
%\newline\newline
{\textbf{Final expression.}} 
Putting everything together, working in the $\Lambda=1$ frame, we have the genus-two contribution $\mathcal{W}^{(2)}(\tilde{\ell})$
%\begin{equation}
 %  \mathcal{W}(\tilde{\ell}) \approx   \frac{1}{4\sqrt{2\pi} \tilde{\kappa}}\left(\tilde{\ell}^{-5/2}+2\tilde{\ell}^{-3/2}\right) e^{-2\tilde{\ell}} +        \frac{\tilde{\kappa}}{24 \sqrt{2 \pi }} \left(\tilde{\ell}^{1/2}+2 \tilde{\ell}^{3/2}\right) e^{-2\tilde{\ell}} + \frac{\tilde{\kappa}^3}{2880\sqrt{2\pi}}\left(20\tilde{\ell}^{9/2}+58 \tilde{\ell}^{7/2}+84\tilde{\ell}^{5/2}+70\tilde{\ell}^{3/2}+35 \right) e^{-2\tilde{\ell}} + O\left(\kappa ^5\right)~.
%\end{equation}
\begin{equation}
\mathcal{W}^{(2)}(\tilde{\ell}) = \frac{\tilde{\kappa}^3}{1474560}\left(5\tilde{\ell}^{\frac{9}{2}}+58 \tilde{\ell}^{\frac{7}{2}}+336\tilde{\ell}^{\frac{5}{2}}+1120\tilde{\ell}^{\frac{3}{2}}+2240\tilde{\ell}^{\frac{1}{2}} \right) e^{-\frac{\tilde{\ell}}{2}}~.
\end{equation}
One can also express the above as  a simple linear combination of Bessel functions. 

\subsection{Higher-genus examples using heat kernel method}\label{Heat kernel calc 2}

Below we present some higher-genus examples that are computed using the Euclidean heat kernel method.
\newline\newline
{\textbf{Genus-three.}} To compute genus-three, we resort to a small Euclidean time heat kernel expansion. We systematically compute an expansion in small $\ell$, and keep the $\mathcal{O}(\kappa^5)$ term, while exponentiating the $\mathcal{O}(\kappa^0)$ part of the potential. We find 
%\begin{multline}
%\mathcal{W}^{(3)}(\tilde{\ell}) = \frac{\tilde{\kappa}^5}{1698693120} \left(6272000\tilde{\ell}^{\frac{1}{2}}+3136000\tilde{\ell}^{\frac{3}{2}}+940800\tilde{\ell}^{\frac{5}{2}}+185280\tilde{\ell}^{\frac{7}{2}}+25440\tilde{\ell}^{\frac{9}{2}}+2448\tilde{\ell}^{\frac{11}{2}} \\
 %+154\tilde{\ell}^{\frac{13}{2}}+5\tilde{\ell}^{\frac{15}{2}} \right) e^{-\frac{\tilde{\ell}}{2}}~.
%\end{multline}
\begin{multline}\label{Genus three wavefunction}
\mathcal{W}^{(3)}(\tilde{\ell}) =  \frac{\tilde{\kappa}^5}{1698693120} \left(6272000\tilde{\ell}^{\frac{1}{2}}+3136000\tilde{\ell}^{\frac{3}{2}}+940800\tilde{\ell}^{\frac{5}{2}}+185280\tilde{\ell}^{\frac{7}{2}} \right. \\
 \left. +25440\tilde{\ell}^{\frac{9}{2}}   +2448\tilde{\ell}^{\frac{11}{2}} +154\tilde{\ell}^{\frac{13}{2}}+5\tilde{\ell}^{\frac{15}{2}} \right) e^{-\frac{\tilde{\ell}}{2}}~.
\end{multline}
Our result (\ref{Genus three wavefunction}) further agrees with the inverse Laplace transform of eqn. (B.1) in \cite{Johnson:2026twg} for $g=3$ once we appropriately rescale the parameters.
\newline\newline
{\textbf{Genus-four.}} As for the genus-three case, again implementing the small Euclidean time heat kernel expansion can be used to compute genus-four. We find
\begin{multline}\label{Genus four wavefunction}
\mathcal{W}^{(4)}(\tilde{\ell}) = \frac{\tilde{\kappa}^7}{13045963161600}\left(318938726400\tilde{\ell}^{\frac{1}{2}} +159469363200\tilde{\ell}^{\frac{3}{2}}+47840808960\tilde{\ell}^{\frac{5}{2}}\right. \\
\left.+9731706880\tilde{\ell}^{\frac{7}{2}} 
 +1448652800\tilde{\ell}^{\frac{9}{2}}+163806720\tilde{\ell}^{\frac{11}{2}}+14232960\tilde{\ell}^{\frac{13}{2}}+942144\tilde{\ell}^{\frac{15}{2}} \right. \\
\left. +45792\tilde{\ell}^{\frac{17}{2}}+1490\tilde{\ell}^{\frac{19}{2}}+25\tilde{\ell}^{\frac{21}{2}}\right) e^{-\frac{\tilde{\ell}}{2}}~.
\end{multline}
We have not found an expression in the literature to compare this to.
\newline\newline
{\textbf{Genus-five.}} Once again, implementing the small Euclidean time heat kernel expansion can be used to compute genus-five. We find
\begin{multline}\label{Genus five wavefunction}
\mathcal{W}^{(5)}(\tilde{\ell}) = \frac{\kappa^9}{5009649854054400}\left(1577726115840000\tilde{\ell}^{\frac{1}{2}}+788863057920000\tilde{\ell}^{\frac{3}{2}}\right.\\
\left.+236658917376000\tilde{\ell}^{\frac{5} {2}}+48844154060800\tilde{\ell}^{\frac{7}{2}}+7517868646400\tilde{\ell}^{\frac{9}{2}}+902117314560\tilde{\ell}^{\frac{11}{2}}\right.\\
\left.+86476974080\tilde{\ell}^{\frac{13}{2}}+6702976000\tilde{\ell}^{\frac{15}{2}}
+420833280\tilde{\ell}^{\frac{17}{2}}+21203840\tilde{\ell}^{\frac{19}{2}}+837184\tilde{\ell}^{\frac{21}{2}}\right.\\
\left.+24672\tilde{\ell}^{\frac{23}{2}}+490\tilde{\ell}^{\frac{25}{2}}+5\tilde{\ell}^{\frac{27}{2}}\right) e^{-\frac{\tilde{\ell}}{2}}~.
\end{multline}
We have again not found an expression in the literature to compare this to.
\newline\newline
{\textbf{Genus-six.}} The small Euclidean time heat kernel expansion can be used to compute genus-six. We find
\begin{multline}\label{Genus six wavefunction}
\mathcal{W}^{(6)}(\tilde{\ell}) = \frac{\kappa^{11}}{57711166318706688000} \left(385712561000546304000\tilde{\ell}^{\frac{1}{2}}\right.\\
\left.+192856280500273152000\tilde{\ell}^{\frac{3}{2}}+57856884150081945600\tilde{\ell}^{\frac{5}{2}}+12039498290875596800\tilde{\ell}^{\frac{7}{2}}\right.\\
\left.+1887114563289088000\tilde{\ell}^{\frac{9}{2}}+233542501888819200\tilde{\ell}^{\frac{11}{2}}+23481506988441600\tilde{\ell}^{\frac{13}{2}}\right.\\
\left.+1952819999907840\tilde{\ell}^{\frac{15}{2}}+135731200389120\tilde{\ell}^{\frac{17}{2}}+7918197862400\tilde{\ell}^{\frac{19}{2}}+387106263040\tilde{\ell}^{\frac{21}{2}}\right.\\
\left.+15731857920\tilde{\ell}^{\frac{23}{2}}+522652992\tilde{\ell}^{\frac{25}{2}}+13775136\tilde{\ell}^{\frac{27}{2}}+272880\tilde{\ell}^{\frac{29}{2}}+3650\tilde{\ell}^{\frac{31}{2}}+25\tilde{\ell}^{\frac{33}{2}}\right)e^{-\frac{\tilde{\ell}}{2}}~.
\end{multline}
As for genus-five we have not found an expression in the literature to compare this to.

\section{Topological gravity as a solvable model}\label{topgrav}

In this appendix, we provide an example of a completely solvable model. This is known as topological gravity, and it is associated to the continuum limit of a purely Gaussian matrix model. Details about this model can be fond in the review \cite{Ginsparg:1993is}. An exposition of the model in a language closer to ours is found in \cite{Gaiotto:2003yb}. The model can be viewed as a  two-dimensional conformal field theory with $c=-2$ coupled to quantum gravity. 

\subsection{String equation, and all genus $\mathcal{W}(\ell)$}

The corresponding string equation is given by
\begin{equation}
\mathcal{P}_\kappa(z) + z = 0~.
\end{equation}
The model is simple because $\mathcal{P}_\kappa(z)$ is independent of $\kappa$. The quantum mechanical problem we must solve is governed by the Hamiltonian
\begin{equation}
H_\kappa = -\frac{\kappa^2}{2}  \partial_z^2 + \frac{1}{2}z~,
\end{equation}
whose eigenfunctions are the Airy functions. The Euclidean heat kernel for the above Hamiltonian can be calculated exactly, and one finds
\begin{equation}
K(z_1,z_2;\ell) = \frac{\exp \left(\frac{\ell^3}{96}\kappa ^2 -\frac{(z_1-z_2)^2}{2 \kappa ^2 \ell}-\frac{\ell}{4}  (z_1+z_2)\right)}{\sqrt{2 \pi  \kappa ^2 \ell}}~.
\end{equation}
The diagonal components are directly read off to be
\begin{equation}
K(z_0,z_0;\ell) =\frac{e^{-\frac{\ell z_0}{2}+\frac{\kappa ^2 \ell^3}{96}}}{\sqrt{2 \pi \kappa ^2 \ell}}~.
\end{equation}
We can thus compute the arbitrary genus macroscopic loop operator %({\color{blue}fix units})
\begin{equation}\label{Wtop}
\mathcal{W}_\kappa(\tilde{\ell}) = \int_1^\infty \dd z_0 K(z_0,z_0;\tilde{\ell}) = \frac{2e^{-\tilde{\ell}/2}}{\sqrt{2 \pi \tilde{\kappa} ^2}} \frac{e^{\frac{\tilde{\kappa} ^2 \tilde{\ell}^3}{96}}}{\tilde{\ell}^{\frac{3}{2}}}~.
\end{equation}
Though very simple, the above calculation exemplifies how perturbative effects in $\kappa$ can resum into a dominant behaviour at large $\tilde{\ell}$. In this case, though normaliseable at large $\tilde{\ell}$ for each order in the small-$\kappa$ expansion, viewed as a Wheeler-DeWitt wavefunction, $\mathcal{W}(\tilde{\ell})$ is badly non-normaliseable at any finite $\kappa$. The origin of the non-normaliseability is the unboundedness of the quantum mechanical linear potential.
\newline\newline
It is also instructive to consider the Laplace transform of (\ref{Wtop}), at each order in the genus expansion. One finds
\begin{equation}
{\mathcal{W}}^{(h)}(\mu_B) =   \frac{\kappa ^{2 h-1}}{12^{h}\sqrt{\pi } h!}\Gamma \left(3 h-\frac{1}{2}\right) (1+2 \mu_B)^{\frac{1}{2}-3 h}~.
%{\mathcal{W}}^{(h)}(\mu_B) =  \sqrt{\frac{2}{\pi }} \tilde{\kappa} ^{2 h-1} \Gamma \left(3 h-\frac{1}{2}\right)  (1-\mu_B)^{\frac{1}{2}-3 h}~.
\end{equation}
We see that the non-analyticity in $\mu_B$ dictates the power of $\tilde{\ell}$ in the genus expansion.
%\subsection{Matrix model treatment, briefly}
\newline\newline
As a brief final comment, we can also obtain the above result from a Gaussian matrix model. Specifically, we need to compute the Gaussian integral over $N\times N$ Hermitean matrices, M, 
\begin{equation}
\mathcal{W}_N(\ell) = \frac{1}{\mathcal{W}_N(0) }\int \dd M e^{- \frac{1}{2g^2} \text{Tr} M^2} \text{Tr} \,e^{-\ell M}~,
\end{equation}
where we have normalised by the undecorated matrix integral. This can be computed using the method of orthogonal polynomials, which are Hermite polynomials for the Gaussian matrix integral. We leave it as an exercise to the reader to show that
\begin{equation}
\mathcal{W}_N(\ell) =  e^{\frac{{g}^2\ell^2}{2}} L^{(1)}_{N-1} \left(-g^2 \ell^2\right)~,
\end{equation}
where $L^{(\alpha)}_{N-1} (z)$ denotes a generalised Laguerre polynomial. To make contact with the double scaling limit, we have to  rescale the integral such that it picks up contributions from the edge of the single cut semi-circle eigenvalue Wigner distribution. The range of the distribution in the planar approximation is given by $\lambda \in 2\sqrt{N}g(-1,1)$. The details of this are worked out in Chapter 9  of \cite{Ginsparg:1993is}, and specifically the worked example in section 9.3 of \cite{Ginsparg:1993is}. Essentially, one shows directly how the orthogonal polynomials, which are Hermite polynomials for the Gaussian matrix model, become Airy functions upon focusing in on the edge of the distribution. Once we have identified the Airy functions, the analysis of the previous subsection follows suit. 

\bibliographystyle{JHEP}
\bibliography{bib}
\end{document}